\documentclass[12pt]{article}
\usepackage{amsmath}
\usepackage{amssymb}
\usepackage{mathtext}
\usepackage{graphicx}

\begin{document}

\section*{LIE ALGEBRAS IN VORTEX DYNAMICS AND \\
CELESTIAL MECHANICS --- IV}
{\bf 1) Classificaton of the algebra of $n$ vortices on a plane\\
2) Solvable problems of vortex dynamics\\
3) Algebraization and reduction in a three-body problem}\footnote{REGULAR AND
CHAOTIC DYNAMICS V.4, No. 1, 1999\\
{\it Received March 22, 1999\\
AMS MSC 76C05}}\bigskip

\begin{centering}
A.\,V.\,BOLSINOV\\
Faculty of Mechanics and Mathematics,\\
Department of Topology and Aplications\\
M.\,V.\,Lomonosov Moscow State University\\
Vorob'ievy Gory, Moscow, Russia, 119899 \\
E-mail: bols@difgeo.math.msu.su\medskip\\
A.\,V.\,BORISOV\\
Faculty of Mechanics and Mathematics,\\
Department of Theoretical Mechanics
Moscow State University\\
Vorob'ievy gory, 119899 Moscow, Russia\\
E-mail: borisov@uni.udm.ru\medskip\\
I.\,S.\,MAMAEV\\
Laboratory of Dynamical Chaos and Non Linearity,\\
Udmurt State University\\
Universitetskaya, 1, Izhevsk, Russia, 426034\\
E-mail: rcd@uni.udm.ru\\
\end{centering}

\begin{abstract}
The work~\cite{BPavlov} introduces a naive description of dynamics  of
 point vortices on a plane in terms of variables of distances and areas
which generate Lie--Poisson structure. Using this approach a
qualitative description of dynamics  of  point vortices on a plane and
a sphere is obtained in the works~\cite{BorisovLebedev,BorisovLeb}.
In this paper we consider more formal constructions of the general problem of~$n$
vortices on a plane and a sphere. The developed methods of algebraization
are also applied to the classical
problem of the reduction in the three-body problem.
\end{abstract}

\section{Classification of the algebra of $n$ vortices on a plane}

\subsection{Vortex algebra and Lie pencils}

Let us introduce coordinates of vortices ${\bf r}_k(x_k,y_k),\;k=1,\dots,n$
by complex variables $z_k= x_k+iy_k,\;k=1,\ldots,n.$
The set ${\bf z}=(z_1,\dots,z_n)$ defines a vector in
 a complex linear space ${\mathbb C}^n$ with an Hermitian form
\begin{equation}
\label{Bol1}
({\bf z,\bf w})_\Gamma=\sum_{i=1}^n \Gamma_i z_i \overline{w}_i,\quad
\Gamma_i\in{\mathbb R}\,,
\end{equation}
where $\Gamma_i$ are vorticities.


The imaginary part
of (\ref{Bol1}) gives the
symplectic form corresponding to  the Poisson
bracket $\{x_i,y_j\}=\frac{\delta_{ij}}{\Gamma_i}$.
 The Hamiltonian~$\mathcal H$ and integrals of motion $P,\,Q,\,I$ of
vortex system can be written as
\begin{gather} \mathcal
H=\frac{1}{16\pi}\sum\limits_{i,j=1}^n \Gamma_i \Gamma_j \ln |z_i-z_j|^2\,,
\label{new1}\\ P+iQ=\sum\limits_{i=1}^n \Gamma_i
z_i=({\bf z},{\bf z}_0)_{\Gamma}\,, \qquad I=\sum\limits_{i=1}^n \Gamma_i
|z_i|^2=({\bf z,z})_{\Gamma}\,, \label{new2} \end{gather}
here
${\bf z}_0=(1,1,\ldots,1)$ is a constant vector.

As relative variables we introduce the squares
of mutual
distances and doubled areas of
triangles spanned on three
vortices (see Fig.~1)
as in~\cite{BPavlov}
\begin{equation}
\label{Bol3/2}
\begin{array}{c}
M_{lm} =({\bf r}_i-{\bf r}_j)^2=(x_{i}-x_{j})^2+(y_{i}-y_{j})^2,\\[8pt]
\Delta_{ijk}=({\bf r}_j-{\bf r}_i)\wedge ({\bf r}_k-{\bf r}_i)=
(x_{j}-x_{i})(y_{k}-y_{i})-(x_{k}-x_{i})(y_{j}-y_{i}).
\end{array}
\end{equation}
 New variables, as it was shown~\cite{BPavlov},
generate a linear bracket, which also
linearly depends on inverse in\-ten\-sities~$1/\Gamma_i$
\begin{align}
\{M_{ij}, M_{kl}\}={}&4\Bigl(\frac{1}{\Gamma_i}\delta_{ik}
-\frac{1}{\Gamma_j}
\delta_{jk}\Bigr)\Delta_{ijl}+4\Bigl(\frac{1}{\Gamma_i}\delta_{il}
-\frac{1}{\Gamma_j}\delta_{jl}\Bigr)\Delta_{ijk},\notag\\
\{M_{ij}, \Delta_{klm}\}={}&
\Bigl(\frac{1}{\Gamma_i}\delta_{ik}-\frac{1}{\Gamma_j}
\delta_{jk}\Bigr)(M_{li}-M_{im}+M_{mj}-M_{jl})+\notag\\
{}\span+\Bigl(\frac{1}{\Gamma_i}\delta_{il}
-\frac{1}{\Gamma_j}\delta_{jl}\Bigr)(M_{mi}-M_{ik}+M_{kj}-M_{jm})+\notag\\
{}\span+\Bigl(\frac{1}{\Gamma_i}\delta_{im}-
\frac{1}{\Gamma_j}\delta_{jm}\Bigr)(M_{ki}-M_{il}+M_{lj}-M_{jk}),\notag\\
\label{Plane7}
\{\Delta_{ijk},\Delta_{lmn}\}={}&
\frac{\delta_{il}}{\Gamma_i}(\Delta_{jkn}-\Delta_{jkm})+
\frac{\delta_{im}}{\Gamma_i}(\Delta_{jkl}-\Delta_{jkn})+{}\\
{}\span+\frac{\delta_{in}}{\Gamma_i}(\Delta_{jkm}-\Delta_{jkl})+
\frac{\delta_{jl}}{\Gamma_j}(\Delta_{ikm}-\Delta_{ikn})+{}\notag\\
{}\span+\frac{\delta_{jm}}{\Gamma_j}(\Delta_{ikn}-\Delta_{ikl})+
\frac{\delta_{jn}}{\Gamma_j}(\Delta_{ikl}-\Delta_{ikm})+{}\notag\\
{}\span+\frac{\delta_{kl}}{\Gamma_k}(\Delta_{ijn}-\Delta_{ijm})+
\frac{\delta_{km}}{\Gamma_k}(\Delta_{ijl}-\Delta_{ijn})+
\frac{\delta_{kn}}{\Gamma_k}(\Delta_{ijm}-\Delta_{ijl}).\notag
\end{align}

Let us set invariant relations corresponding to real motions~\cite{BPavlov}
\begin{equation}
F_{ijkl}=\Delta_{ijk}+\Delta_{ikl}-\Delta_{lij}-\Delta_{ljk}=0\,,
\label{Plane11}
\end{equation}
\vspace{-7mm}
\begin{equation}
F_{ijk}={(2\Delta_{ijk})}^2+M_{ij}^2+M_{jk}^2+
M_{ik}^2-2(M_{ij}M_{jk}+M_{ij}M_{ik}+M_{jk}M_{ik})=0\,,
\label{plane}
\end{equation}
Relations (\ref{Plane11}) mean
that the quadrangle spanned on the vortices $ijkl$ can be constructed
of triangles by two
 ways (Fig.~1). Equations~(\ref{plane}) are Heron formulas
expressing the area of
a triangle  via its sides.

Let us note that for the structure (\ref{Plane7})
 the Jacobi
identity is fulfilled only on the submanifold defined by the first set of
geometrical relations~(\ref{Plane11}).
Below, we consider the
relations~(\ref{Plane11}) to be fulfilled on default.  Thus the
brackets~(\ref{Plane7}) define Lie--Poisson structure, which wee shall call
{\it vortex bracket}.

The bracket (\ref{Plane7}) admits  the linear Casimir function
\cite{BPavlov}
\begin{equation}
\label{Plane10}
D=\sum_{i,j=1}^N   \Gamma_i   \Gamma_j    M_{ij}=2\left(
I\left(\sum\limits_{i=1}^n \Gamma_i\right)-Q^2-P^2\right)\,.
\end{equation}


To define  the real type of the corresponding Lie
algebra we indicate an explicit isomorphism with some Lie
pencil~\cite{Bolsinov}.

{\bf Remark 1.} {\it The compactness conditions of the
real form~(\ref{Plane7}), depending on intensities,
imply that all mutual distances are bounded from above
at any moment of time.}

We consider a space of skew Hermitian  $n\times n$ matrices  and
subspace $L$  in it generated by matrices of the form:
\begin{equation}
\label{Bol2} \begin{array}{c} \Delta_{lmk} = \begin{matrix} \mbox{}\\  l  \\
\mbox{}  \\ m \\ \mbox{} \\ k \\ \mbox{}\end{matrix} \left( \arraycolsep=3pt
\begin{array}{ccccccc} \ddots & &      &     &      &     &   \\[-3pt] & 0
&      &  1  &      & -1  & \\[-3pt] &      &\ddots&     &      &     &
       \\[-3pt] &-1    & &  0  &      &  1  &   \\[-3pt] &      &      &
       &\ddots&     & \\[-3pt] & 1    &      & -1  &      &  0  &
       \\[-3pt] &      & &     &      &     &\ddots \end{array} \right),
       \\ M_{lm} = \begin{matrix} \mbox{}\\  l  \\ \mbox{}  \\ m \\ \mbox{}
\end{matrix} \left( \arraycolsep=8pt
\begin{array}{ccccc} \ddots &      &
&     &     \\[-3pt] & i    &      & -i  &     \\[-3pt] &      &\ddots&
       &     \\[-3pt] &-i    &      &  i  &     \\[-3pt] &      &      &
       &\ddots \end{array} \right).  \end{array} \end{equation}

Let us describe the main properties of this subspace which can be verified by a
straightforward calculation.

{\bf Proposition 1.}
\begin{itemize}
\item[1)]
{\it The $L$ is  a Lie subalgebra,}

\item[2)] {\it This subalgebra is isomorphic to $u(n-1)$,}

\item[3)]
{\it The center of this subalgebra is generated by the element} 
$\sum\limits_{m,l} M_{ml}$,

\item[4)]  {\it The elements $\Delta_{lms}$ generate a
subalgebra in $L$ isomorphic to $so(n-1)$,}

\item[5)] {\it The
following relation holds
for any} $k,l,m,s$:
$$ \Delta_{klm} + \Delta_{kms} +
\Delta_{lks} + \Delta_{mls} = 0, $$

\item[6)] {\it Commutation relations in
$L$ coincide with the vortex bracket
(\ref{Plane7}) for the case when all intensities are the same (and equal
to unity).  }\end{itemize}

Therefore, (\ref{Bol2}) determines
an  $n$-dimensional (unitary) linear representation of Lie algebra
 ${u(n-1)}$. There are no such irreducible representations,
that is why it splits into a sum of the standard
 ${n-1}$-dimentional representation and one-dimensional
trivial representation.  This splitting is organized
as follows.

{\bf Proposition 2.}
{\it The representation space $V = \Bbb C^n$ splits
 into a direct sum of invariant subspaces
$V = V_1 \oplus V_2$, where $V_1 ={\mathbb C}^{n-1}$ is given by
$z_1 + z_2 + \dots + z_n =  0$ and a one-dimensional subspace $V_2  =
{\mathbb C}$
is spanned on the vector  ${\bf z}_0=(1,1,\dots, 1)$.}

Let us turn to the case of arbitrary intensities  $\Gamma_1,\dots,\Gamma_n.$
Consider  the Lie pencil on the algebra of
skew Hermitian $n\times n$ matrices generated by commutators

\begin{equation}
\label{eqqq*} [X, Y]_{\Gamma^{-1}} = X\Gamma^{-1} Y - Y\Gamma^{-1} X,
\end{equation} where $\Gamma^{-1}$ is a real diagonal matrix of the form
\begin{equation} \label{eq46_3} \Gamma^{-1} = \left( \begin{matrix} \frac
{1}{\Gamma_1}   &    &    &    \\ & \frac {1}{\Gamma_2} & & \\ & & \ddots
& \\ & &        & \frac {1}{\Gamma_n} \end{matrix} \right).  \end{equation}

The remarkable fact is that the subalgebra $L$ is closed
with respect to the commutator  $[\cdot,\cdot]_{\Gamma^{-1}}$.
Thus the  family of commutators
~\eqref{eqqq*} generates some Lie pencil on $L$. Moreover,
restricting the commutator $[\cdot,\cdot]_{\Gamma^{-1}}$ on $L$  we
obtain the Lie algebra isomorphic to the vortex
algebra~$L_\Gamma$ corresponding to intensities $\Gamma_1, \dots,
\Gamma_n$. Thus one can deduce a symmetric isomorphism between the family
of vortex algebras and a rather simple Lie pencil.

Using this construction we shall describe
properties of vortex algebras.

{\bf Proposition 3.}
{\it For positive $\Gamma_i$ the vortex algebra is isomorphic
to~${u(n-1)}$.}

{\it Proof.}

The corresponding isomorphism is constructed as follows. From the beginning
let all intensities be equal to unity. All matrices~\eqref{Bol2}
are skew Hermitian and satisfy the following property~$A{\bf z}_0=0$,
where~${\bf z}_0$ is  the vector with coordinates
~$(1,\,1,\,\ldots,\,1)$. In other words, this vector is invariant
under  the vortex algebra.
Therefore its orthogonal completment, i.~e., the hyperplane~$V_1=
\{{\bf z} \mid\sum z_i = 0\}$ is also invariant.

We  consider all skew Hermitian matrices possessing this
property (that is, mapping ${\bf z}_0$ into zero) 
They form  the subalgebra~$u(n-1)$.  Thus the
vortex algebra in
case of equal intensities is embedded into~$u(n-1)$.
But their dimensions coincide, therefore the
algebras also coincide.

This can be shown more explicitly by choosing 
another basis in space $\Bbb C^n$.  Consider a basis $\bf e_1, \bf e_2, \dots
, \bf e_n$ where $\bf e_1, \dots , \bf e_{n-1}$ is  an 
orthonormal
basis in  the hyperplane $V_1$ (see Proposition~2) and
${\bf e}_n = (1/\sqrt {n},\dots, 1/\sqrt {n})$.  Rewriting
matrices~(\ref{Bol2}) from the vortex algebra in the new basis one can see
that they take the form
\begin{equation}
\label{new3_j}
\left( \begin{matrix} A_{n-1} & \begin{matrix} 0 \\
\vdots \\ 0 \end{matrix} \\ \begin{matrix} 0 & \hdots  & 0 \end{matrix} & 0
\end{matrix} \right),
\end{equation}
where $A_{n-1}$ is a skew Hermitian
 $(n-1)\times (n-1)$ matrix. In this basis the isomorphism of the vortex algebra with $u(n-1)$ is evident.
A defect  of this basis is that this basis can't be made symmetric with
respect to all vortices.

Let us generalize these arguments to the case of arbitrary
intensities.
Transition  to the new basis in~${\mathbb C}^n$ defines the  conjugation of
matrices
in algebra~(\ref{Bol2}) of the form $A = C  A'  C^{-1}$,  where  $C$ is
the transformation matrix.
(In our case  we can assume it to be real and
orthogonal ~$C^{-1}=C^T$).  With this substitution the commutator of the
vortex algebra transforms to the following form:
$$ \begin{aligned}
A\Gamma^{-1} B - B \Gamma^{-1} A& = CA' C^{-1} \Gamma^{-1}
C B' C^{-1} - CB' C^{-1}
\Gamma^{-1} C A' C^{-1} =\\
\mbox{}&= C [A', B']_{\Gamma '} C^{-1},
\end{aligned}
$$
where
$$
\Gamma'= C^{-1} \Gamma^{-1} C =
\left( \begin{matrix}
\Gamma'_{n-1} & \begin{matrix} \gamma_{1n} \\ \vdots \\ \gamma_{n-1, n}
\end{matrix} \\
\begin{matrix} \gamma_{1n} & \hdots & \gamma_{n-1, n} \end{matrix} & \gamma_{n,n}
\end{matrix} \right).
$$

Here $\Gamma'_{n-1}$ is the symmetric $(n-1)\times (n-1)$ matrix
corresponding to the restriction of the form~$\Gamma^{-1}$ onto 
the hyperplane $V_1$. This follows from the relation $C^{-1} = C^T$.

Taking into account that the last row
and column of matrices $A$ and $B$ vanish one finds that the vortex
commutator transforms to the form
$$ [A'_{n-1} ,
B'_{n-1}]_{\Gamma'_{n-1}}.  $$

Let us note that the matrix~$\Gamma'_{n-1}$ is positively defined and real,
therefore one can  take the root~$(\Gamma'_{n-1})^{1/2}$.
Then the substitution
$$
A'_{n-1} =
(\Gamma'_{n-1})^{-1/2}A''_{n-1}(\Gamma'_{n-1})^{-1/2}
$$
reduces the commutator
$[A'_{n-1} , B'_{n-1}]_{\Gamma'_{n-1}}$ to the standard one. The matrices
remain skew Hermitian. This argument shows also that the
vortex pencil under consideration is a subpencil of the standard pencil on
 the space of skew Hermitian matrices.$\blacksquare$


The main result following from this construction is that the properties of
 the vortex Lie algebra corresponding to parameters
$\Gamma_1, \dots, \Gamma_n$ are completely determined by properties of
 the bilinear form~$\Gamma_{n-1}'$. This form is the
restriction of the
 form~$\Gamma^{-1}$  onto the subspace
 $V_1 = \{ z_1 + \dots + z_N =
0\}$ (simply by the signature of this restriction).

{\bf Remark 2.}
{\it Using the method of proof of Proposition 3 it is not
difficult to show that in general case the vortex algebra is isomorphic
to the algebra~$u(p,q)$ with~$p+q=n-1$.}

Let us find conditions for which   the 
algebra $L_\Gamma$ is compact, i.~e.,
isomorphic to  the Lie algebra $u(n-1)$.
The necessary and sufficient condition is that $\Gamma_{n-1}'$
is a form of fixed sign. In this case there are the
 following possibilities (we suppose that intensities are finite
and differ from zero).

\begin{itemize}
\item[1)] The $\Gamma^{-1}$ is a form of fixed
sign, i.~e., all $\Gamma_i$
are simultaneously positive  or negative.
\item[2)] The form  $\Gamma^{-1}$ has the signature $(n-1,  1)$,   all
$\Gamma_i$  except for  one are positive.
Indeed the condition of positive definiteness 
of the restriction on~$V_1$
 requires that the form $\Gamma^{-1}$ is negative defined onto 
a one-dimensional orthogonal  complement to the  $V_1$ with respect to~$\Gamma$ 
This condition is easy to find if one notices that the orthogonal
 complement to  $V_1$ in the sense of~(\ref{eq46_3}) is a one
dimensional subspace spanned on the vector $v_\Gamma =
(\Gamma_1, \dots , \Gamma_n)$. It has the form
$$
(v_\Gamma , v_\Gamma )_{\Gamma^{-1}} = \sum \Gamma_i < 0.
$$
\item[3)] Similarly, in the case of the signature $(1, n-1)$.
\end{itemize}

Finally we have the following

{\bf Proposition 4.}
{\it The vortex Lie algebra $L_\Gamma$ is compact only in the following
cases:}
\begin{itemize}
\item[1)] {\it All intensities have the same sign.
\item[2)] All intensities except for one are positive and
$\sum\limits_{i=1}^n \Gamma_i < 0$.
\item[3)] All intensities except for one are negative and
$\sum\limits_{i=1}^n \Gamma_i > 0$.}
\end{itemize}

{\bf Remark 3.} {\it
This proposition coincides with the one which was proved in
 the case of three vortices \cite{BorisovLebedev}.}

If  the vortex algebra is not
``semi-simple", the conditions are defined similarly.
It happens iff the form
$\Gamma'_{n-1}$ is degenerate on $V_1$. From linear algebra it follows that
this condition equals to the requirement that the
orthogonal complement  to $V_1$ lies in
 $V_{1}$. It means that
$v_\Gamma \in V_1$, i.~e.
$$
\Gamma_1 + \ldots + \Gamma_n = 0.
$$

Let us note one more point. The subspace $\Delta \subset L$ spanned on vectors
 $\Delta_{lmn}$ is a subalgebra for any algebra of the pencil.
Its compactness conditions are the same as  those
for the whole $L_\Gamma$, i.~e., the form $\Gamma_{n-1}'$ should be
positive definite. Thus the vortex algebra $L_\Gamma$ is
compact iff the subal\-gebra~$\Delta$ is compact.

\subsection{Reduction with symmetries and singular orbits}

Now we explain the origin
the nature of Lie pencils connected with the vortex algebra and describe
the (singular)
symplectic orbits of the algebra~(\ref{Plane7}) corresponding to real motions.
Consider a transition to relative variables~(\ref{Bol3/2}) from the point
of view of reduction with symmetries \cite{MarsdenW}.

Hamiltonian function~\eqref{new1} and equations of motion of vortices are
invariant under the action of the group~$E(2)$. This action can be represented in
the form
\begin{equation}
\label{eqq_dop1}
g\colon{\bf z}\to e^{i\varphi}{\bf z}+a{\bf z}_0\,,\quad g\in E(2)\,,
\end{equation}
where~${\bf z}_0=(1,\,\ldots,\,1),\;a=a_x+ia_y$.
Parameters~$\varphi,\,a_x,\,a_y$ define
the translation and rotation corresponding to the element~$g\in E(2)$.

The action~\eqref{eqq_dop1} is the {\em non-Hamiltonian} group action~\cite{Arnold}.
Indeed integrals of motion~\eqref{new2} cor\-res\-pon\-ding to translations
 $P,\,Q$ and
to rotations $I$ form a Poisson structure that differs from Lie--Poisson structure
of the algebra~$e(2)$ by a constant, cocycle.
Obviously this cocycle is unremovable and standard reduction
with the momentum is impossible~\cite{Arnold, ArnoldGivental, MarsdenW}.
To execute the reduction in algebraic form we use the momentum map in some other
way.

Consider the action of linear transformation
preserving the
form~$(\cdot,\cdot)_{\Gamma}$~\eqref{Bol1}.
The corresponding linear operators
 (matrices) form the group which is isomorphic to~$U(p,q)$, $p+q=n$ and satisfies
relations
$$
X\Gamma X^+=\Gamma\,,\qquad \Gamma=\mbox{diag}(\Gamma_1,\,\ldots,\,\Gamma_n)\,.
$$
Operators of the corresponding Lie algebra~$A\in u(p,q)$ are defined by
\begin{equation}
\label{eqq_dop2}
A\Gamma+\Gamma A^+=0\,.
\end{equation}

Since $\Gamma^+=\Gamma$ from~\eqref{eqq_dop2},
the matrix~$A\Gamma$ is skew Hermitian. After the
substitution $\phi(A)=A\Gamma$ the standard commutator transforms into the
commutator
$[\cdot,\cdot]_{\Gamma^{-1}}$~\eqref{eqqq*}:
$$
\phi([A,B])=[\phi(A),\phi(B)]_{\Gamma^{-1}}\,.
$$
The advantage of this approach is that for any~$\Gamma_i$ the
algebra~\eqref{eqq_dop2} is represented on the same space of skew Hermitian
matrices. However, instead of the standard commutator it is necessary to
use~$[\cdot,\cdot]_{\Gamma^{-1}}$. In this case a family 
of Poisson
brackets~$\{\cdot,\cdot\}_{\Gamma^{-1}}$ appears on
the corresponding coalgebra~$u^*(p,q)$.

Linear vector fields corresponding to operators~\eqref{eqq_dop2} in complex form
can be written as
$$
V_A=A{\bf z}=\phi(A)\Gamma^{-1}{\bf z}\,.
$$
These fields  are Hamiltonian~\cite{ArnoldGivental} with
\begin{equation}
\label{eqq_dop3}
H_A=\frac{i}{2}(A{\bf z},{\bf z})_\Gamma=\frac i2\overline{{\bf z}}^TA
\Gamma{\bf z}=\frac
i2(\phi(A){\bf z},{\bf z})=\frac{i}{2}\overline{{\bf z}}^T\phi(A){\bf z}\,.
\end{equation}
The bracket~$\{\cdot,\cdot\}_{\Gamma^{-1}}$ of quadratic
Hamiltonian functions is
\begin{equation}
\label{eqq_dop4}
\{H_{\phi(A)}({\bf z}),H_{\phi(B)}({\bf z})\}_{\Gamma^{-1}}=\frac {i}{2}\bigl(
[\phi(A),\phi(B)]_{\Gamma^{-1}}{\bf z},{\bf z}\bigr),
\end{equation}
and it corresponds to the commutator~$[\cdot,\cdot]_{\Gamma^{-1}}$.

We can describe the momentum map~$\mu({\bf z})$
explicitly using
 the standard  identification~$u(n)$ and~$u^*(n)$ with the help of the form
$\langle A,B\rangle=\mbox{Tr}AB$
$$
\mbox{Tr}\mu({\bf z})\phi(A)=\mbox{Tr}\frac {i}{2}\overline{{\bf z}}^T\phi(A){\bf z}
=\mbox{Tr}\left(\frac {i}{2}{\bf z}\overline{{\bf z}}^T\right)\phi(A).
$$
Thus
\begin{equation}
\label{eqq_dop5}
\mu({\bf z})=\frac {i}{2}{\bf z}\overline{{\bf
z}}^T=\frac{i}{2}\begin{pmatrix}
z_1\overline {z}_1&\cdots&z_1\overline {z}_n\\
\vdots&\ddots&\vdots\\
z_n\overline {z}_1&\cdots&z_n\overline {z}_n
\end{pmatrix}.
\end{equation}
The formula~\eqref{eqq_dop5} defines a map onto a singular symplectic orbit
of the algebra~$u(p,q)$ with the
com\-mu\-ta\-tor~$[\cdot,\cdot]_{\Gamma^{-1}}$.  Now the
integral~(\ref{new2})~$I=\sum\Gamma_i z_i\overline{z}_i$ is a Casimir
function, therefore the reduction with it is carried out. In case of all
positive (negative) intensities the orbit is topologically homeomorphic
to~${\mathbb C}P^{n-1}$, because under the map~(\ref{eqq_dop5}) all points
of the form ~$e^{i\varphi}{\bf z}$ (orbits  of the rotation group
action~\eqref{eq46_3}) stick together.

Let us carry out the reduction with the remained integrals~$P,Q$~\eqref{new2}
on the reduced space of matrices~(\ref{eqq_dop5}).
Due to non-commutativity of~$P,Q$ we can reduce only one degree of
freedom~\cite{ArnoldKozlovNei}.
The constant vector field corresponding to the translation in the
direction~$a=a_x+ia_y$
is generated by the
linear Hamiltonian
$$
H_a=({\bf z},\Gamma a{\bf z}_0)=({\bf z},a{\bf z}_0)_\Gamma\,.
$$
It's easy to show that Hamiltonians $H_A$~\eqref{eqq_dop3} commuting with~$H_a$
are generated by matrices~$A$ for which
$$
A\Gamma {\bf z}_0 =\phi(A){\bf z}_0=0\,.
$$
So~$\phi(A)$ belong to the subspace~$L$ defined in the previous section
 (matrices~(\ref{eqq_dop5}) don't belong to this space).

Now it is
easy to see that  squares of mutual distances and areas~(\ref{Bol3/2})
also admit
a natural representation of the form
\begin{align*}
M_{ij}(x,y)&=|z_i-z_j|^2=i(M_{ij}{\bf z},{\bf z})\,,\\
\Delta_{ijk}(x,y)&=i(\Delta_{ijk}{\bf z},{\bf z})\,,
\end{align*}
where~$M_{ij},\,\Delta_{ijk}$ are matrices~\eqref{Bol2}. The
 momentum map~$\mu({\bf z})$ into the corresponding algebra~$u(p',q')$, $p'+q'=n-1$
has the form
\begin{equation}
\label{eq_dop_*}
\mu({\bf z})=\frac{i}{2}\left({\bf z}-\frac{({\bf z},{\bf z}_0)\Gamma{\bf
z}_0}{\sum\Gamma_i}\right) \left(\overline{\bf
z}-\frac{({\bf z},{\bf z}_0)\Gamma\overline{\bf
z}_0}{\sum\Gamma_i}\right)^T\,.
\end{equation}
Matrices~(\ref{eq_dop_*}) satisfy  the relation~$\mu({\bf z}){\bf
z}_0=0$, i.~e.,
they belong to the subspace~$L$ defined in the previous
section. In compact case the appropriate
orbit is homeomorphic
to~${\mathbb C}P^{n-2}$
and corresponds to the reduced phase space (after reducing two degrees of
freedom).
The rank of matrices~(\ref{eq_dop_*}) is equal to the unit. This is their
characteristic property.

{\bf Remark 4.}
{\it
The structure and compactness conditions of the reduced phase space can
be found by investigation (by methods of analytical geometry) of a common
level surface of the first integrals~(\ref{new2}). It is
interesting that
the algebraic approach also allows to solve this pure geometrical
problem.}

\subsection{Symplectic coordinates}
Let us describe an algorithm of construction of symplectic coordinates
for the reduced system of~$n$ vortices in the
case of compact
algebra~$(u(n-1))$.  The method of reduction of the vortex algebra (for
this case) to the standard representation has been  described above. In
this case elements of the algebra are represented by skew Hermitian
$(n-1)\times(n-1)$ matrices with an ordinary matrix commutator:
\begin{equation}
\label{eqq_dp1} A=\begin{pmatrix}
 ih_1&x_{12}+iy_{12}&\cdots&x_{1,n-1}+iy_{1,n-1}\\
-x_{12}+iy_{12}&ih_2&\cdots&x_{2,n-1}+iy_{1,n-1}\\
\hdotsfor{4}\\
-x_{1,n-1}+iy_{1,n-1}&-x_{2,n-1}+iy_{2,n-1}&\cdots&ih_{n-1}.
\end{pmatrix}
\end{equation}
Let us define a Poisson map~${\mathbb R}^{2(n-1)}\to u^*(n-1)$ onto
a singular orbit~$u^*(n-1)$  by formulas
\begin{equation}
\label{eqq_dp2}
\begin{aligned}
h_i&=p_i\,,\\
x_{ij}&=\sqrt{h_ih_j}\sin(\varphi_i-\varphi_j)\,,\\
y_{ij}&=\sqrt{h_ih_j}\cos(\varphi_i-\varphi_j)\,,
\end{aligned}
\qquad i,\,j=1,\,\ldots,\,n-1\,.
\end{equation}
The canonical bracket~$\{\varphi_i,p_j\}=\delta_{ij}$ turns in the standard
Lie--Poisson bracket on~$u^*(n-1)$. The orbit via variables~$p_i,\,\varphi_i$
is written as
$$
p_1+p_2+\cdots+p_{n-1}=C=const\,.
$$
This orbit is formed by matrices of rank~1
($2\times 2$~minors
equal  zero).
To introduce symplectic coordinates we carry out a symplectic transformation
\begin{equation}
\label{new**}
\begin{aligned}
r_0&=p_1+p_2+\cdots+p_{n-1}\,,\quad&s_0&=\varphi_1\,,\\
r_1&=p_2+\cdots+p_{n-1},\quad&s_1&=-\varphi_1+\varphi_2\,,\\
&\vdots&{}&\vdots\\
r_{n-2}&=p_{n-2}\,,\quad&s_{n-2}&=-\varphi_{n-1}+\varphi_{n-2}\,.
\end{aligned}
\end{equation}
It is easy to see that the map~\eqref{eqq_dp2} doesn't depend on~$s_0$,
therefore $s_1,\, \ldots,\,s_{n-2}$, $r_1,\,\ldots,\,r_{n-2}$ define
symplectic coordinates of~$2(n-2)$-dimensional orbit $\mathbb CP^{n-2}$.

{\bf Remark 5.}
{\it (Co)algebra~$(u^*)u(n-1)$
admits a linear
change of variables~$h_i\mapsto h_i+ \lambda D,\;
D=\sum\limits_{i=1}^{n-1}h_i,\;\lambda=const$ preserving commutator  
relations. In this connection,
 the orbits given by matrices~$F\in u(n)$ of rank~1 map into 
orbits for which the
matrix~$(A-\lambda D)$ has rank 1.}

{\it The orbit given by Heron relations~(\ref{plane}) (in compact case) after
the 
transition from variables~$M,\triangle$ to standard variables of
algebra~$u(n-1)$ turns out among the above indicated orbits for
some~$\lambda$.}

\subsection{Canonical coordinates of a reduced four-vortex system.
Poincar\'e section}
Using  the above described algorithm we shall point out
canonical coordinates explicitly for a four-vortex problem of equal intensities.

Consider the
coordinates on the algebra~$u(3)$ corresponding to the matrix representation
of the form~(\ref{eqq_dp1})
\begin{equation}
\label{vort1}
A_3=\begin{pmatrix}
ih_1&x_3+iy_3&-x_2+iy_2\\
-x_3+iy_3&ih_2&x_1+iy_1\\
x_2+iy_2&-x_1+iy_1&ih_3,\\
\end{pmatrix}
\end{equation}
In accordance with the above arguments these coordinates
are expressed linearly via areas and
squares of distances
\begin{equation}
\label{vort2}
x_i=\sum_r\delta^r_{(x_i)}\Delta_r{(z)}\mbox{, }y_i=\sum_{k<l}m_{(y_i)}^{kl}
M_{kl}(z)\mbox{, } h_i=\sum_{k<l}m_{(h_i)}^{kl}M_{kl}(z),\\
 i=1,\dots ,3\,.
\end{equation}

To define coefficients~$\delta_{(x_i)}^k,m_{(y_i)}^{kl}
m_{(h_i)}^{kl}$ let us use the Proposition~3 and a matrix realization
(\ref{Bol2}) of elements~$\Delta_r,M_{kl}$,
where~$\Delta_1=\Delta_{234}$, $\Delta_2=-\Delta_{134}$,
$\Delta_3=\Delta_{124}$, $\Delta_4=-\Delta_{123}$. We choose the
orthogonal matrix~$C$, which reduces matrices of the vortex algebra~\eqref{vort2}
by the transformation~$C^TAC$ to the form~(\ref{new3_j}) as:
\begin{equation}
\label{vort3}
C=\begin{pmatrix}
-\frac{1}{2}&-\frac{1}{2}&-\frac{1}{2}&\frac{1}{2}\\
\frac{1}{2}&\frac{1}{2}&-\frac{1}{2}&\frac{1}{2}\\
\frac{1}{2}&-\frac{1}{2}&\frac{1}{2}&\frac{1}{2}\\
-\frac{1}{2}&\frac{1}{2}&\frac{1}{2}&\frac{1}{2}
\end{pmatrix}.
\end{equation}
Let us set one of coordinates~$h_i,x_i,y_i$ in the obtained matrix~$A_3$ is equal
to~1
and the others are equal to~0 and solve the system of linear equations.
Thus we  find corresponding
coefficients~$\delta_{(x_i)}^k,m_{(y_i)}^{kl},m_{(h_i)}^{kl}$. In this case
\begin{equation}
\label{vort4}
\begin{aligned}
x_1&=\frac{1}{4}(-\Delta_1+\Delta_2+\Delta_3-\Delta_4)\mbox{, }&
x_2&=\frac{1}{4}(-\Delta_1+\Delta_2-\Delta_3+\Delta_4)\,,\\
x_3&=\frac{1}{4}(-\Delta_1-\Delta_2+\Delta_3+\Delta_4)\mbox{, }&
y_1&=\frac{1}{4}(M_{14}-M_{23})\,,\\
y_2&=\frac{1}{4}(M_{13}-M_{24})\mbox{, }&
y_3&=\frac{1}{4}(M_{12}-M_{34})\,,\\
h_1&=\frac{1}{4}(M_{12}+M_{34}+M_{13}+M_{24}-M_{14}-M_{23})\,,\span\span\\
h_2&=\frac{1}{4}(M_{12}+M_{34}-M_{13}-M_{24}+M_{14}+M_{23})\,,\span\span\\
h_3&=\frac{1}{4}(-M_{12}-M_{34}+M_{13}+M_{24}+M_{14}+M_{23})\,.\span\span
\end{aligned}
\end{equation}
Denoting canonical coordinates~(\ref{new**}) as~$(g,G,h,H)$, we have
\begin{equation}
\label{vort5}
\begin{array}{c}
\begin{aligned}
x_1&=-\sqrt{(H-G)G}\sin g\mbox{, }&
x_2&=\sqrt{(D-H)G}\sin (h+g)\,,\\[3pt]
x_3&=-\sqrt{(D-H)(H-G)}\sin h\mbox{, }&
y_1&=\sqrt{(H-G)G}\cos g\,,\\[3pt]
y_2&=\sqrt{(D-H)G}\cos (h+g)\mbox{, }&
y_3&=\sqrt{(D-H)(H-G)}\cos h\,,
\end{aligned}
\medskip\\
h_1=D-H,\qquad h_2=H-G,\qquad h_3=G,
\end{array}
\end{equation}
where~$D=\frac{1}{4}\sum_{kl}M_{kl}$ is a constant of the Casimir function~$u(3)$,
in this connection  $0<G<H<D$.

The Hamiltonian in new variables can be represented in the form
\begin{equation}
\label{vort6}
\begin{aligned}
{\cal H}={}&\frac{1}{4\pi}\Bigl\{\ln\bigl((D-G)^2-4(D-H)(H-G)\cos^2h\bigr)+{}\\
{}&+\ln\bigl((D-H+G)^2-4(D-H)G\cos^2(h+g)\bigr)+{}\\
{}&+\ln\bigl(H^2-4(H-G)\cos^2g\bigr)\Bigr\}\,.
\end{aligned}
\end{equation}

Let us construct the Poincar\'e map on a plane~$(g,G)$ for different values
of the energy~$E ={\cal H}(h,H,g,G)$. Surfaces of energy function in
figures~2--4
for fixed~${g=g_0}$, ${G=G_0}$, $f(h,H)={\cal
H}(h,H,g,G)$ show the complex structure of the isoenergetic surface.

We define an intersecting plane by the equation~$H=H_*$. The value~$H_*$
should be chosen so that~$\forall g_0,G_0<H_*$ equation~$f(h,H_*)=E$
has the unique solution with positive (negative)
deri\-va\-tion~$\frac{dH}{dt}$. As is obvious from the presented
Fig.~2--4, that it is fulfilled not for all
$H_*$ (analogous conditions for the intersecting plane
$h=h_*$ are practically never fulfilled).

The maximal possible energy~$E_T=-4\ln2\simeq -2.77\ldots$ corresponds
to the Tompson configuration for~$D=1$, in this connection~$H=G=\frac{1}{2}$.
In this case
the phase portrait on a plane~$(g,G)$ consists of the only
line~$G=\frac{1}{2}$. Phase portraits for the energies~$ E<E_T$ are
represented in Fig.~5--10.
Regions where the motion is impossible~($f(h,H_*)=E$ has no solutions)
corresponds to painted regions in Fig.~5--10.
Figures show that with decreasing of the energy the stochastic layer increases
first
occupying all plane (Fig.~7, 8). Then it decreases
and remains only near unstable solutions and
sepa\-ra\-tri\-ces~(Fig.~9, 10).
In the limit~$E\to -\infty$ one of pairs
of vortices merges and we obtain an integrable problem that is
the system of three vortices.

\subsection{Lax--Heisenberg representation}
As a result of the reduction we  can  write equations of motion
on the orbit of the 
coadjoint representation of the Lie
algebra $u(p,q)$.  This orbit is singular and consists of matrices of the
form $$ L = \frac{i}{2} {\bf z \bar \bf z}^{\text {T}}\,, $$ where $\sum z_i =
0$ (in a system connected with the center of vorticity).

According to a general principle  (because of the simplicity
of the 
corresponding Lie algebra~\cite{BolsBor})
we can rewrite equations in Lax--Heisenberg form:
$$
\frac {dL}{dt} = [L, A]\,,
$$
where $A=dH(L)$ is the 
differential of the 
Hamiltonian of the form
$$
H(L) = \sum \Gamma_i \Gamma_j \log M_{ij}(L)\,.
$$

Here $M_{ij}$ is interpreted as an element of
the 
Lie algebra
$u(n-1)$ and $M_{ij}(L)$~ is a standard pairing between the algebra and
coalgebra.  In our case the  identification  $M_{ij}(L)=\pm \mbox{Tr} M_{ij} L =
|z_i - z_j|^2$ is fulfilled.

The explicit form for the differential of
the 
Hamiltonian is given by the formula:
$$
dH(L) = \sum \Gamma_i \Gamma_j \frac {1}{ M_{ij}(L)} M_{ij}\,,
$$
i.~e.,  $dH(L)$ is a linear combination of matrices $M_{ij}$.
Thus the expression for the matrix ~$A$ has the form
$$
A = dH(L) = i \left( \begin{matrix}
\sum_j \frac {\Gamma_1 \Gamma_j}{|z_1 - z_j|^2} &
-\frac {\Gamma_1 \Gamma_2}{|z_1 - z_2|^2} & \ldots &
-\frac {\Gamma_1 \Gamma_n}{|z_1 - z_n|^2} \\[5pt]
-\frac {\Gamma_2 \Gamma_1}{|z_2 - z_1|^2} &
\sum_j \frac {\Gamma_2 \Gamma_j}{|z_2 - z_j|^2} & \ldots &
-\frac {\Gamma_2 \Gamma_n}{|z_2 - z_n|^2} \\[5pt]
\vdots & \vdots & \ddots & \vdots \\
-\frac {\Gamma_n \Gamma_1}{|z_n - z_1|^2} &
\ldots &
-\frac {\Gamma_n \Gamma_{n-1}}{|z_n - z_{n-1}|^2} &
\sum_j \frac {\Gamma_n \Gamma_j}{|z_2 - z_j|^2}
\end{matrix} \right).
$$

If all intensities coincide with each other and
 equal to 1, the matrix
$A$ is simplified:
$$
A = i \left( \begin{matrix}
\sum_j \frac {1}{|z_1 - z_j|^2} &
-\frac {1}{|z_1 - z_2|^2} & \ldots &
-\frac {1}{|z_1 - z_n|^2} \\[5pt]
-\frac {1}{|z_2 - z_1|^2} &
\sum_j \frac {1}{|z_2 - z_j|^2} & \ldots &
-\frac {1}{|z_2 - z_n|^2} \\[5pt]
\vdots & \vdots & \ddots & \vdots \\
-\frac {1}{|z_n - z_1|^2} &
\ldots &
-\frac {1}{|z_n - z_{n-1}|^2} &
\sum_j \frac {1}{|z_2 - z_j|^2}
\end{matrix} \right).
$$

In case of different intensities the additional procedure is required
to reduce the commutator to the standard form.

\subsection{Stationary configurations}
In terms of $L-A$ pair the stationary configuration is naturally described.
The stationary condition is equivalent to
the fact that the matrices~$L$ and~$A$ commute:
$[{L},{A}]=0$. In our case it means that
$$
\frac {i}{2} A{\bf z}\bar {\bf z}^{\text{T}} = \frac {i}{2} {\bf z} \bar {\bf z}^{\text{T}}
A =
-\frac {i}{2} {\bf z} (\bar {A{\bf z}})^{\text{T}} .
$$

Denoting ${\boldsymbol b}=A{\bf z}$, one has
${\boldsymbol b}\bar {\bf z}^{\text{T}} = - {\bf z} \bar {\boldsymbol b}^{\text{T}}$.
It can be shown that it is possible only if   $\boldsymbol b =
i\lambda {\bf z}$, $\lambda \in \Bbb R$.
It means that ${\bf z}$ is an eigenvector of a matrix
  $A$ (with an imaginary eigenvalue).

As a result we obtain a rather natural set of relations
$$
 \left( \begin{matrix}
\sum_j \frac {1}{|z_1 - z_j|^2} &
-\frac {1}{|z_1 - z_2|^2} & \ldots &
-\frac {1}{|z_1 - z_n|^2} \\
-\frac {1}{|z_2 - z_1|^2} &
\sum_j \frac {1}{|z_2 - z_j|^2} & \ldots &
-\frac {1}{|z_2 - z_n|^2} \\
\vdots & \vdots & \ddots & \vdots \\
-\frac {1}{|z_n - z_1|^2} &
\ldots &
-\frac {1}{|z_n - z_{n-1}|^2} &
\sum_j \frac {1}{|z_2 - z_j|^2}
\end{matrix} \right)
\left( \begin{matrix} z_1 \\ z_2 \\ \vdots \\ z_n \end{matrix} \right) = \lambda
\left( \begin{matrix} z_1 \\ z_2 \\ \vdots \\ z_n \end{matrix} \right).
$$
This can be rewritten in a simpler form
\begin{equation}
\label{bolsin*}
\sum_j \frac {1}{z_k - z_j} = \lambda \bar z_k\mbox{, where } k=1, \dots, n.
\end{equation}

Equations~(\ref{bolsin*}) can be obtained directly from
the 
equations of motion under the condition that every point rotates around
the origin with the same velocity~$\lambda$.

Stationary configurations have been studied in several works
(see~\cite{ArefH,Campbell,Melesh}). Apparently new results can be obtained
with the help of the following arguments.

\begin{itemize}
\item[1.]  Stationary configurations can be interpreted as eigenvectors of
a matrix $A$.
\item[2.] Stationary configurations can be interpreted as singular points
of Hamiltonian on an orbit (which is on $\mathbb{C}P^{n-1}$). As a Hamiltonian
one can take a function $\tilde H =
\prod_{k\ne j} |z_k - z_j|^2$. It is a positive function which vanishes
on a submanifold of the  ``collapse".  \end{itemize}

Using these arguments and  by the investigation of a commutativity condition
$[{L,A}]=0$, it would be interesting to give a theoretical explication of
configurations from the ``Los-Alamos" catalogue.
For these configurations stable states of rotation
realize on  concentric circles
(``atomic covers" by Kelvin) (see Fig.~11), where for the system of
11 vortices two
possibilities~$11=2+9$ or $11=3+8$)~\cite{Aref2}  are pointed out.
As a rule these configurations possess the same type of symmetry (a rotational one
or have a plane of symmetry). Recently in a short note in~``Nature"~\cite{ArefV}
non-symmetric stationary configurations have been indicated for the vortex
system of equal intensities.

Theoretically the simplest problem is to determine a  number of collinear
configurations in depen\-dence on the ratio of intensities. In celestial mechanics
(for positive masses~$m_i$) the answer is given by the Multon theorem.
In accordance  with this theorem there is the only (rotating) collinear
configuration
for any permutation of masses~$m_1, \dots, m_n$ $(m_i>0)$
(the proof of this theorem is contained in~\cite{Smeil}). For the three-vortex
system
the number of collinear configurations depends on the type of the algebra
defined by Poisson brackets  \cite{BorisovLebedev,BorisovLeb}.
Apparently there is such a connection in general case
 (for the system of~$n$ vortices).

 \section{Solvable problems of vortex dynamics}
In this section we consider in
parallel the problems of dynamics of point vortices on a plane and sphere
\cite{Bog2, BPavlov}. The position of vortices on a sphere in the "absolute" space
is defined in cartesian~${\bf r}_i (x_i, y_i, z_i) $ or spatial polar
coordinates~$ (\theta_i, \varphi_i, \, i=1.. N) $
\[
x_i=R\sin \theta_i \cos \varphi_i \,,\qquad
y_i=R\sin \theta_i \sin \varphi_i \,,\qquad
z_i=R\cos \theta_i \,,
\]
where $R $ is the radius of a sphere. Let us remind (see \cite{BPavlov}), that
the Hamiltonian function and Poisson bracket are defined by relations
\[
\begin{aligned}
{}\span
H = -\frac{1}{8 \pi} \sum\limits _{i < j} \Gamma_i \Gamma_j \ln 4R^2 \sin^2
\frac{\theta _{ij}}{2} \,, \\
{}\span
\{\varphi_i, \cos \theta_j \} =\frac{1}{R^2
\Gamma_i} \delta _{ij} \,, \qquad
i, j=1, \ldots, N \,,
\end{aligned}
\]
where $ \theta _{ij} $ is the angle between $i $-th and $j $-th vortices, and $
\Gamma_i $ is a vortex intensity. For relative variables on a sphere the same
labels --- $M _{ij} $, $ \Delta _{ijk} $ are used, however, their meaning is
different, than for a plane~\cite{BPavlov}. So $M _{ij} $ is a length of a
chord connecting the vortices, and $ \Delta _{ijk} $ is equal to the ratio of
volume of a tetrahedron spanned on three vortices and the center of a sphere
to~$\frac{1}{4} R $.
\begin{equation} \label{c1}
\begin{aligned}
 M_{ij}=&{}(x_i-x_j)^2+(y_i-y_j)^2+(z_i-z_j)^2\,,\\
\Delta_{ijk}=&{}\frac{1}{R}({\bf r} _i{, \bf r} _j\times{\bf r} _k) \,.
\end{aligned}
\end{equation}

\subsection{The particular case of the problem of $n$ vortices, the reduction
to the problem of ${(n-1)}$ vortices}

There is a special case of the problem of $n$ vortices on a plane and a sphere,
for that case the problem can be reduced to the system of $(n-1)$ vortices with
the same algebra of Poisson brackets, but reduced Hamiltonian function.
The procedure of a reduction in Hamiltonian exposition corresponds to
restriction of the system on a Poisson submanifold \cite{Fomenko2}, which is
defined in this case by involution conditions  of integrals of motion.
For vortices on a plane the integrals are given by relations~(\ref{new2}),
necessary conditions of reduction accept the form
\begin{equation}
\label{Plane25.1}
\sum _{i=1} ^N\Gamma_i=0 \,,\quad Q=P=0 \,.
\end{equation}
For a sphere integrals of motion are obtained similarly:
\begin{equation}
\label{Sp5}
\begin{gathered}
F_1 = R\sum_{i=1}^N\Gamma_{i}\sin\theta_{i}\cos\varphi_{i}\,,\quad
F_2=R\sum_{i=1}^N\Gamma_{i}\sin\theta_{i}\sin\varphi_{i}\,,\quad
F_3 = R\sum _{i=1} ^N\Gamma _{i} \cos\theta _{i}\,,
\end{gathered}
\end{equation}
and conditions of reduction
\begin{equation}
\label{Plane25.2}
F _{1} =F _{2} =F _{3} =0\,.
\end{equation}

{\bf Remark 6.}{\it
The equations (\ref{Plane25.1}) have the following geometrical meaning: each
vortex is situated at the center of  vorticity of all remaining vortices.
Really, expressing for example $\displaystyle\Gamma_N=-\sum_{i=1}^{N-
1}\Gamma_i$ from (\ref{Plane25.1}), we have found that the absolute coordinates
of the $N-$th vortex are defined by expressions}
$$
{\bf r}_N=\frac{\sum\limits_{i=1}^{N-1}\Gamma_i{\bf r} _i}{\sum\limits _{i=1}
^{N-1} \Gamma_i}\,.
$$
{\it The geometrical meaning of relations (\ref{Plane25.2}) on a sphere is a little
different, it means, that the center of  vorticity of a system of $n$ vortices
coincides with the geometrical center of a sphere.}

{\it The mentioned below reduction happens because there is a possibility of
calculation of positions of~$n$ vortices basing on positions of~$(n-1)$
vortices.}

For a determination of additional invariant relations for relative variables we
shall use identities~\cite{Melesh}
\begin{equation}
\label{Plane25.3}
\frac {1}{2}\sum _{k=1} ^N\Gamma_k (M _{jk} -M _{ik}) =
P (x_i-x_j) +Q (y_i-y_j)
\end{equation}
for plane which are fulfilled only under condition of $\displaystyle
\smash{\sum_{i=1}^N}\Gamma_i=0$, that is also valid for the case of sphere~\cite{BPavlov}
\begin{equation}
\label{Plane25.4}
\frac {1}{2}\sum _{k=1} ^N\Gamma_k (M _{jk} -M _{ik}) =
R\bigl(F_{1}(x_i-x_j)+F_{2}(y_i-y_j)+F_{3}(z_i-z_j)\bigr)\,.
\end{equation}

Using conditions (\ref{Plane25.1}), (\ref{Plane25.2}), we have obtained
$C_N^2$ invariant relations for $M_{ij}$
\begin{equation}
\label{Plane25.5}
\sum _{k=1} ^N\Gamma_k (M _{jk} -M _{ik}) =0\,.
\end{equation}
These relations we shall supplement by relations for $\Delta_{ijk}$, which also
follow from (\ref{Plane25.1}), (\ref{Plane25.2}). It is possible to show by a
straightforward calculation that the complete set of these relations determines a
Poisson submanifold with the structure~(\ref{Plane7}) and its generalization for a
sphere \cite{BPavlov}.

Using representation for $M_{ij}$ in absolute coordinates (\ref{Bol3/2})
and~(\ref{c1}), it is simple to verify that among the equations~(\ref
{Plane25.5}) only $n-1$ are linearly independent. Using these equations, we can
express squares of distances from one of the vortices ($n$th vortex) to all
the others~$M_{kn},\; k=1,\ldots,n-1$ via mutual distances between
$n-1$ vortices~$M_{ij}, \; i,j=1,\ldots,n-1$.
Then substituting these expressions into initial Hamiltonian, we shall obtain a
system with a bracket of the problem of~$n-1$ vortices and reduced
Hamiltonian function.

For the case of four vortices the explicit formulas of squares of distances
from the first three vortices up to the fourth vortex $M_{14}, M_{24}, M_{34}$
have
a form
\begin{equation}
\label{Plane25.6}
\hookrightarrow
 M_{14}=\frac{\Gamma_3^2M_{13}+\Gamma_2^2M_{12}+ \Gamma_2\Gamma_3 (M _{12} -M _
{23} +M _{13})}{
( \Gamma_1 +\Gamma_2 +\Gamma_3) ^2}\,,
\end{equation}
the reduced Hamiltonian is obtained by substitution (\ref{Plane25.6}) into
initial Hamiltonian (\ref{new1}).

The system of three vortices is integrable (independently on a Hamiltonian
given on al\-geb\-ra~(\ref{Plane7})), the described case corresponds to a special
case of  integrability of problem of four vortices (origina\-ting in work by
Kirchhoff~\cite{Kirh}).
The phase portraits of this system at various intensities are presented in the
previous work \cite{BorisovLeb}.

{\bf Remark 7.} {\it
The indicated special cases of integrability correspond to a situation at
which one of integrals reaches the extreme value. It is obvious, that in this
case the system necessarily has additional invariant relations. For integrable
systems that situation produces an additional degeneration. Delone case for
Kovalevskaya top is one of examples. In this case the Kovalevskaya integral, which is
the sum of two full squares will vanish, and the two-dimensional tori degenerate
into one-dimensional torus(periodic and asymptotic solutions).}

\subsection{Particular solutions in the problem of four vortices}
General equations of motion of vortices~\cite{BPavlov} with some restrictions on
intensities~$\Gamma_i$ allow a finite group of symmetries, the elements of the
group are the permutations and reflections in some planes. Such discrete
symmetries do not produce the general integrals of motion and do not allow
reducing the order of a system. However, presence of these symmetries is
connected to invariant submanifolds. The solution on that submanifolds can, as a
rule, be obtained in quadratures~\cite{Melesh}.

Let us consider two problems of dynamics of four vortices on a plane and a sphere
possessing two types of symmetry --- central symmetry and reflection symmetry
(for a plane, central symmetry and axial symmetry).

\noindent{\bf a. A centrally symmetric solution at $\bf {D=0}$.} Equations
of motion
of four vortices on sphere (flat case is obtained with the passage to the
limit~$R\to\infty$) with condition~$\Gamma_1 =\Gamma_3,$ $ \Gamma_2 =\Gamma_4$
allow invariant relations
\begin{equation}
\label{Plane34}
\begin{array}{cc}
M _{14} =M _{23} =M _{14} =M_1 \,,\quad &M_{12}=M_{34}=M_3\,,\\
\Delta_{234}=\Delta_{124}=\Delta_{1}\,,\quad &\Delta_{213}= \Delta _{314} =
\Delta _{2} \,.
\end{array}
\end{equation}
(The relations~(\ref{Plane34}) don't define the Poisson submanifold).


The equations~(\ref{Plane34}) have the following geometrical meaning: a
centrally symmetric configuration of vortices (the parallelogram), keeps this
symmetry in all moments (see Fig.~12).

The analysis of a centrally symmetric solution in absolute variables through
explicit quadratures is carried out by D.\,N.\,Gorya\-chev~\cite{Goryachev1}
(see also~\cite{ArefPoint}). However, he has not made clear the qualitative
properties of motion and has introduced the very confusing classification. Let us
introduce the qualitative analysis of relative motion.

The equations describing evolution of sides  $M_1, M_3$ and diagonals
$M_{13}=M_2 $, $M_{24}=M_4$ of the parallelogram, have a form
\begin{equation}
\label{Plane35}
\begin{array}{l}
\dot M_{1}=4\Gamma_1\Delta_2\left(\frac{1}{M_2}-\frac{1}{M_3}\right)
+ 4\Gamma_2\Delta_1\left(\frac{1}{M_4}-\frac{1}{M_3}\right)\,,\\[10pt] \dot
M_{3}=4\Gamma_1\Delta_2\left(\frac{1}{M_1}-\frac{1}{M_2}\right)
+ 4\Gamma_2\Delta_1\left(\frac{1}{M_1}-\frac{1}{M_4}\right)\,,\\[10pt] \dot
M_{2}=8\Gamma_2\Delta_2\left(\frac{1}{M_3}-\frac{1}{M_1}\right)\,,
\\ [10pt] \dot
 M_{4}=8\Gamma_1\Delta_1\left(\frac{1}{M_3}-\frac{1}{M_1}\right)\,.
\end{array}
\end{equation}

For a sphere (for a plane $R\to \infty $) the geometrical relations between
$M, \Delta $ can be written as
\begin{equation}
\label{cen3}
\begin{array}{l}
\Delta_1\left(4R^2-M_2\right)+\Delta_2\left(4R^2-M_1-M_3\right)=0\,,\\[6pt]
\Delta_1\left(4R^2-M_1-M_3\right)+\Delta_2\left(4R^2-M_4\right)=0\,.\\[6pt]
\end{array}
\end{equation}
System~(\ref{cen3}) is solvable with respect to $ \Delta_1, \Delta_2$ under
condition of
\begin{equation}
\label{cen4}
2(M_1+M_3)-(M_2+M_4)+\frac{1}{4R^2}\left(M_2M_4-(M_1+M_3)^2\right)=0\,. \end
{equation}
The linear integral of the equations~(\ref{Plane35}) corresponding to the Casimir
function (\ref{Plane10}) has a form
\begin{equation}
\label{cen5}
 D=2\Gamma_1\Gamma_2(M_1+M_3)+\Gamma_1^2M_2+\Gamma_2^2M_4\,.
\end{equation}

With the help of relations~(\ref{cen3}) and regularization of time, the
system~(\ref{Plane35}) can be reduced to two inhomogeneous equations describing
evolution of sides of the parallelogram~$M_1,\,M_3$.

For simplicity we consider the limited case~$D=0$, that is  also a necessary
condition of a collapse~\cite{NovikovSed}. Geometrical interpretations on a
plane and a sphere are a little different, therefore we shall consider these cases
separately.\medskip

1) P\,l\,a\,n\,e.

Relations (\ref{cen3}),(\ref{cen4}) for a plane ($R\to\infty $) have a form
\begin{equation}
\label{cen6}
\begin{array}{c}
\Delta_1 =-\Delta_2 =\Delta \,, \\ [6pt]
2 (M_1+M_3) - (M_2+M_4) =0 \,.
\end{array}
\end{equation}
Taking into account, that in (\ref{cen5}) $D=0$, we found the equation,
describing a trajectory of system on plane $M_1, M_3 $ (sides of a
parallelogram)
\begin{equation}
\label{cen8}
\frac{dM_1}{dM_3} = -\frac
{M_1\left(2\Gamma_1\Gamma_2(M_1+M_3)+(\Gamma_1^2+\Gamma_2^2)M_3\right)}
{M_3\left(2\Gamma_1\Gamma_2(M_1+M_3)+(\Gamma_1^2+\Gamma_2^2)M_1\right)}\,.
\end{equation}
The solution of this equation has a form
\begin{equation}
\label{cen9}
\begin{array}{c}
\left(M_1-M_3\right)^2=\left(M_1+M_3\right)^2 -
C\left(M_1+M_3\right)^{-2\alpha}\,,\\[6pt]
\alpha=\frac{\Gamma_1^2+\Gamma_2^2}{4\Gamma_1\Gamma_2}\,,\qquad C =const \,.
\end{array}
\end{equation}
(Index $\alpha<0$, because with $D=0$ intensities $\Gamma_1, \Gamma_2$ have
different signs).

In a quadrant $M_1 > 0, \, M_2 > 0 $ (which corresponds to physical area)
three types of trajectories~(\ref{cen9}) are possible depending on $\alpha$:
\begin{itemize}
\item [$ 1^\circ.$] $-1<\alpha<0$~--- all trajectories are closed, started from
the origin of coordinates, tangenting axes $OM_1 $ $OM_2 $ (see. Fig.~13, a),
a) (inhomogeneous collapse).
\item [$ 2^\circ.$] $\alpha <-1$~--- all trajectory are curves asymptotically
approaching to coordinate axes (see  Fig.~13,b).
\item [$ 3^\circ.$] $\alpha =-1$~--- all trajectories are straight lines moving
through the origin of coordinates (see  Fig.~13,c) (homogeneous
collapse)~\cite{NovikovSed}.
\end{itemize}

The physical area on a plane~$ M_1,M_3$ is defined by the part of a positive
quadrant~$(M_1>0$, ${M_3>0)}$, for which the inequality~$\Delta^2>0$ is fulfilled.
When the trajectory reached a boundary~${\Delta^2=0}$, the sign in the
equation~(\ref{cen8}) should be changed (reflection) that corresponds to the
same trajectory passing in the opposite direction. It is easy to show, that the
equation~$\Delta^2=0$ is defined by two straight lines on a plane~$M_1,M_3$,
which are situated inside a quadrant ~ $ M_1 > 0, \, M_3 > 0 $ for any
intensities~$\Gamma_1, \Gamma_2$. Hence, except for a case~$3^\circ$ vortices
are moving in the bounded area without collisions and scattering.
In the system in case~$3^\circ$ the homogeneous collapse of all vortices or
homogeneous scattering happen.

For the problem on a plane each point of the trajectory on the plane~$M_1,M_3$
correspond to two configurations of vortices, which differ only by
permutation:~$1\leftrightarrow 3$, or~$2\leftrightarrow4$ (see
Fig.~12).

{\bf Remark 8.} {\it
The condition of a homogeneous collapse ($\alpha=-1$) from the analysis of
motion in absolute variables is obtained in~\cite{NovikovSed}. As
result of
quasi-homogeneity of the equations the condition of a homogeneous collapse can
be obtained by investigation of existence conditions  of solutions of the
form $M_i=C_i t^{\xi},\, C_i=const$ of the system.}

{\bf Remark 9.} {\it
For a plane the system~(\ref{Plane35}) with a regularization}
$$
d\tau =\frac{4\Delta}{M_1M_2M_3M_4} \, dt
$$
{\it can be reduced to a homogeneous Hamiltonian system}
\begin{align}
\frac{dM_1}{d\tau}&=\Gamma_1M_1M_4(M_3-M_2)+\Gamma_2M_1M_2(M_4-M_3)\,,\\[-3pt]
\frac{dM_3}{d\tau}&=\Gamma_1M_3M_4(M_2-M_1)+\Gamma_2M_2M_3(M_1-M_3)\,,\\[-3pt]
\frac{dM_2}{d\tau}&=2\Gamma_2M_2M_4(M_1-M_3)\,,\\[-3pt]
\frac{dM_4}{d\tau}&=2\Gamma_1M_2M_4(M_3-M_1)
\end{align}
{\it with a Poisson bracket of the form}
\begin{align}
\{M_1,M_2\}&=\frac{1}{\Gamma_1}M_1M_2M_3M_4\,,\quad&\{M_1,M_3\}&=
\frac12\left(\frac{1}{\Gamma_1}-\frac{1}{\Gamma_2}\right)M_1M_2M_3M_4\,,\notag\\[-3pt
] \label{cencen}
\{M_2,M_3\}&=\frac{1}{\Gamma_1}M_1M_2M_3M_4\,,\quad&\{M_1,M_4\}&= -\frac1
{\Gamma_2} M_1M_2M_3M_4\,, \\[-3pt]
 \{M_2,M_4\}&=0,\quad&\{M_3,M_4\}&=
\frac1{\Gamma_2} M_1M_2M_3M_4\notag
\end{align}
{\it and Hamiltonian}
$$
H=2\Gamma_1\Gamma_2\ln M_1 +\Gamma_1^2\ln M_2+2\Gamma_1\Gamma_2\ln M_3 +
\Gamma_2^2\ln M_4\,.
$$

{\it The rank of a bracket~(\ref{cencen}) is equal to two, hence dividing the
bracket on~$M_1M_2M_3M_4$, it can be reduced to a constant without violation of
the Jacobi identity.}
\medskip

2) S\,p\,h\,e\,r\,e.

For a centrally symmetric configuration on a sphere we also consider a
projection of trajectories to a plane $M_1,M_3$ (see Fig.~14).
The form of physical area defined by inequalities $\Delta_i^2>0,\,i=1,2$
varies. Moreover, there are two differences from a case of plane connected with the
nonlinearity of the equation (\ref{cen4}). At first, each point $M_1,M_3$
correspond to two solutions of the equation (\ref{cen4}), and therefore two
various (spatial) parallelograms with the given sides on a sphere, these
parallelograms are not reduced to each other by permutation~$1\leftrightarrow
3\;(2\leftrightarrow 4)$. Secondly, the system~(\ref{Plane35}) is not reduced
to two quasi-homogeneous equations.

Let us express~$\Delta_1$ and $\Delta_2$ from~(\ref{cen3}) with preservation of
homogeneity
\begin{equation}
\label{regul1}
\Delta_1 = (M_4-M_1-M_3) \Delta \,,\qquad
\Delta_2 = (M_2-M_1-M_3) \Delta \,.
\end{equation}
Also we shall make change of the time
$$
 dt=\smash{\frac{4\Delta}{M_1M_3}}\,d\tau\,.
$$
So we obtained the system describing evolution of
variables~$M_1$,~$M_2$,~$M_3$,~$M_4$ \vspace{-2mm}
\begin{equation}
\label{evolut_*}
\begin{aligned}
 M'_1&=M_1\left(\Gamma_1\frac{(M_2-M_1-M_3)(M_3-M_2)}{M_2}
 +\Gamma_2
 \frac{(M_4-M_1-M_3)(M_3-M_4)}{M_4}\right)\,,\\
 M'_3&=M_3\left(\Gamma_1\frac{(M_2-M_1-M_3)(M_2-M_1)}{M_2}
 +\Gamma_2
 \frac{(M_4-M_1-M_3)(M_4-M_1)}{M_4}\right)\,,\\
M'_2&=2\Gamma_2(M_2-M_1-M_3)(M_1-M_3)\,,\\
M'_4&=2\Gamma_1(M_4-M_1-M_3)(M_1-M_3)\,,
\end{aligned}
\end{equation}
here primes denote derivation on new time $\tau$.

Let us introduce a linear change of variables reducing system (\ref{evolut_*})
to two differential equations, which has the most simple form.

Let us choose variables $x, y, M, N $:
\begin{equation}
\label{perem_*}
\begin{aligned}
 x&=\Gamma_1M_2+\Gamma_2M_4\,,\qquad &  y&=\Gamma_1^2M_2-\Gamma_2^2M_4\,,\\
M&=M_1+M_3\,,\qquad &  N&=M_1-M_3\,.
\end{aligned}
\end{equation}
With the help of a relation (\ref{cen5}) where $D=0$ and (\ref{cen4}) which have
a form
\begin{equation}
\label{1a}
 x=-{\frac{y^2}{16R^2\Gamma_1\Gamma_2(\Gamma_1+\Gamma_2)}}\,,
\end{equation}
we exclude $x, M $ from~(\ref{evolut_*}). In outcome we obtain the equations
of evolution $y, N $\vspace{-1mm}
\begin{align}
\label{eqq}
\frac{dy}{d\tau}=&\,4(\Gamma_1+\Gamma_2)yN\,,\notag\\
\frac{dN}{d\tau}=&-\frac{1}{4(\Gamma_1+\Gamma_2)^3\Gamma_1^2
\Gamma_2^2}\biggl[\frac{2ky-\Gamma_1(\Gamma_1+\Gamma_2)}{\Gamma_2
(ky-\Gamma_1(\Gamma_1+\Gamma_2))}\Bigl(4\Gamma_1^3\Gamma_2^2(\Gamma_1 +
\Gamma_2) ^4N^2-\notag \\
&-y^2(2ky-\Gamma_1^2+\Gamma_2^2)(2ky(\Gamma_1+2\Gamma_2) -
\Gamma_1(\Gamma_1+3\Gamma_2)(\Gamma_1+\Gamma_2))\Bigr)+\notag\\
&+\frac{2ky+\Gamma_2(\Gamma_1+\Gamma_2)}{\Gamma_1
(ky+\Gamma_2(\Gamma_1+\Gamma_2))}\Bigl(4\Gamma_1^2\Gamma_2^3(\Gamma_1 +
\Gamma_2) ^4N^2- \\[-3pt]
&-y^2(2ky-\Gamma_1^2+\Gamma_2^2)(2ky(\Gamma_2+2\Gamma_1)
+\Gamma_2(\Gamma_2+3\Gamma_1)(\Gamma_1+\Gamma_2))\Bigr)\biggr]\,.\notag \end
{align}
hereinafter $k=1/R^2 $.

The projection of trajectory~$y(\tau),\, N(\tau)$ on the plane $M_1,M_3$ is
defined by the formulas $M_1=\frac 12 (M+N),M_3=\frac 12 (M-N)$, where
\begin{equation}
\label{2a}
 M=\frac{\Gamma_2-\Gamma_1}{2\Gamma_1\Gamma_2(\Gamma_1+\Gamma_2)}y+
\frac{y^2}{16R^2\Gamma_1\Gamma_2(\Gamma_1+\Gamma_2)^2}\,,\qquad
(\Gamma_1\Gamma_2 < 0) \,.
\end{equation}
The equation (\ref{2a}) allows real solutions only for
$M\le M_{max}=-\frac{(\Gamma_1-\Gamma_2)^2}{\Gamma_1\Gamma_2}R^2.$ Therefore, in
case of a sphere the physically accessible area on a plane is defined by
inequalities
\begin{gather}
M_1+M_3\le M _{max} \,,
\label{4a}
\\[4pt]
\begin{aligned}
4\Delta_1^2=&\,2(M_1M_4+M_3M_4+M_1M_3)-M_1^2-M_3^2-M_4^2-\frac{1}{R^2}M_1M_3M_
4 \ge 0 \,, \\
4\Delta_2^2=&\,2(M_1M_2+M_3M_2+M_1M_3)-M_1^2-M_3^2-M_2^2-\frac{1}{R^2}M_1M_3M_
2\ge 0 \,.
\end{aligned}
\label{3a}
\end{gather}

In the area (\ref{3a}) the equation (\ref{2a}) has two roots, therefore we have
on a plane $M_1,M_3$ a projection of two different areas of possible positions of
vortices defined by inequalities (\ref{4a}) and equations~(\ref{2a}).
If the areas (\ref{4a}) do not reach a straight line $M_1+M_3=M_{max}$
(Fig.~14 b, c), the point, beginning to move inside one area,
remains there in all moments. When the point reached boundaries ($ \Delta_i=0$),
it is necessary to change a sign of time, and the trajectory is passed in the
opposite direction.
In a case when the straight line $M_1+M_3=M_{max} $ passes through the interior
of areas~(\ref{4a}) (Fig.~14 a) the trajectory, that had reached
that line, passes from one area in another and is described by other solution of
the equation (\ref{2a}). Characteristics of phase portraits in case of a sphere
is the occurrence new motionless points, absent in a flat case.
If the second necessary condition of a collapse is realized
($\Gamma_1^2+\Gamma_2^2=-4\Gamma_1\Gamma_1$), we can see on a Fig.~14
 c), that some of trajectories starting at the origin of coordinates goes
to origin again and do not reach the boundary of area (collinear positions) and
some trajectories goes back to origin only after the reaching of boundary. It
means, that the collapse remains homogeneous only near the origin of
coordinates.

\noindent{\bf b. A reflective-symmetric solution}.
For a system of two cooperating vortex pairs~$\Gamma_1=-\Gamma_4$, ${\Gamma_2=-
\Gamma_3}$ the general system of four vortices on a sphere and on a plane also
allows invariant relations
\begin{equation}
\label{zerc1}
\begin{array}{cc}
M _{12} =M _{34} =M_1\,, \quad &M_{13}=M_{24}=M_{3}\,,\\[5pt] \Delta _{123} =
\Delta _{234} = \Delta_1\,, \quad &\Delta_{124}=\Delta_{134}= \Delta _{2}\,.
\end{array}
\end{equation}
These equations have the following geometrical meaning. The vortices located at
the initial momentum in tops of a trapezoid will form a trapezoid during all time
of motion (see Fig.~15) \cite{Melesh}. The given configuration has a reflection
(axial) symmetry. The equations describing the evolution of sides and diagonals
of a trapezoid have a form
\begin{equation}
\label{zerc2}
\begin{aligned}
\dot M_1&=4\Gamma_2\Delta_1\left(\frac{1}{M_4}-\frac{1}{M_3}\right)
+4\Gamma_1\Delta_2\left(\frac{1}{M_2}-\frac{1}{M_3}\right)\,,\\
\dot M_3&=4\Gamma_2\Delta_1\left(\frac{1}{M_4}-\frac{1}{M_1}\right)
+4\Gamma_2\Delta_2\left(\frac{1}{M_2}-\frac{1}{M_1}\right)\,,\\
\dot M_2&=8\Gamma_2\Delta_2\left(\frac{1}{M_1}-\frac{1}{M_3}\right)\,,\\ \dot
M_4&=8\Gamma_1\Delta_1\left(\frac{1}{M_1}-\frac{1}{M_3}\right)\,. \end
{aligned}
\end{equation}
Here~$ M_2=M_{14}, \; M_{4}=M_{23}$ --- of the bases of a trapezoid.

The geometrical relations between~$M,\,\Delta$ in this case have the identical
form on a plane and sphere:
\begin{equation}
\label{zerc3}
\begin{aligned}
 \Delta_1M_4+\Delta_2(M_3-M_1)&=0\,,\\
 \Delta_1(M_3-M_1)+\Delta_2M_2&=0\,,\\
\end{aligned}
\end{equation}
the condition of their resolvability with respect to $ \Delta_1, \, \Delta_2 $ has a
form
\begin{equation}
\label{zerc4}
M_2M_4- (M_1-M_3) ^2=0 \,.
\end{equation}
The integral of the momentum has a form
\begin{equation}
\label{zerc5}
 D=2\Gamma_1\Gamma_2(M_1-M_3)-\Gamma_1^2M_4-\Gamma_2^2M_2\,.
\end{equation}
From the equations~(\ref{zerc2}--\ref{zerc4}) it follows that
trajectories of a system in space~$ M_1,M_2,M_3,M_4$ coincide for cases
of a plane and sphere. The difference between these problems consists
in form of physical areas defined by inequalities~(\ref{4a}).

With the help of the equations~(\ref{zerc3}) we shall
express~$\Delta_1,\Delta_2 $ by
$$
 \Delta_1=(M_2-M_1+M_3)\Delta,\quad\Delta_2=-(M_4-M_1+M_3)\Delta\,,
$$
and also we shall exclude~$\Delta$ with the help of regularizing substitution of
time
$$
dt =\frac{4\Delta}{M_1M_3} d\tau \,.
$$
To reduce the regularized equation to the system of two equations, we make the
substitution~(\ref{perem_*}). As a result of the existence of an integral of
the moment between the variables~$ x, y $ the following relation is fulfilled
\begin{equation}
\label{zerc6}
\frac{4\Gamma_1\Gamma_2}{\Gamma_1+\Gamma_2}Dx+y^2+\frac{2(\Gamma_1\Gamma_2)}
{\Gamma_1+\Gamma_2}Dy+D^2=0\,.
\end{equation}
Using~(\ref{zerc6}) and equation~(\ref{zerc4}), we find
\begin{equation}
\label{zerc7}
 N=-\frac{y^2}{4\Gamma_1\Gamma_2D}+\frac{D}{4\Gamma_1\Gamma_2}\,.
\end{equation}
With~$ D\ne0 $ from relations~(\ref{zerc5}),~(\ref{zerc6}) and~(\ref{zerc7})
all~$ M_1, M_2, M_3, M_4 $ can be expressed through~$ M, y $. For the
variables~$M, y $ we shall get the equations of the form
\begin{equation}
\label{zerc8}
\begin{array}{c}
\frac{dy}{d\tau}=\frac{(y^2-D^2)(y(\Gamma_1+\Gamma_2)+D(\Gamma_1-\Gamma_ 2))}
{4\Gamma_1\Gamma_2D}\,, \\[12pt]
\begin{aligned}
\frac{dM}{d\tau}=&\,\frac{1}{32D^2\Gamma_1\Gamma_2^3(D-y)}\left[D(\Gamma_2 -
\Gamma_1)-y(\Gamma_2+\Gamma_1)\bigr)\bigl(16D^2\Gamma_1^2\Gamma_2^2M^2 +
\right. \\[5pt]
 &\left.+8(D-y)^2D\Gamma_2^2M-(D^2-y^2)^2\right]-\\[10pt]
&-\frac{1}{32D^2\Gamma_1^3\Gamma_2(D+y)}
\left[D(\Gamma_2-\Gamma_1)-y(\Gamma_2+\Gamma_1))
(16D^2\Gamma_1^2\Gamma_2^2M_2+\right.\\[5pt]
&\left.+8(D+y)^2D\Gamma_1^2M-(D^2-y^2)^2\right]\,.
\end{aligned}
\end{array}
\end{equation}
By analogy with a centrally symmetric solution, we shall consider a projection of
a trajectory to a plane~$ M_1, M_3 $. Besides inequalities~(\ref{4a}) the
physical area is defined by an additional condition (corollary of the
relation~(\ref{zerc7}))
\begin{equation}
\label{zerc9}
D (D-4\Gamma_1\Gamma_2 N) > 0 \,.
\end{equation}
Two various points of a phase space (two solutions of the
equation~(\ref{zerc7})) are projected in each point on a plane~$ M_1, M_3 $,
satisfying to inequalities~(\ref{4a}),~(\ref{zerc9}). It is more evident if we
imagine that two different areas defined by an inequality~(\ref{4a}) are glued
together along a line~$M_1-M_3=N_{0}$. When a point reached a
boundary~$\Delta_i=0$, it is reflected back and moves on the same trajectory,
and at reaching boundary~(\ref{zerc9}) the point passes from one area to
another (see. Fig.~16 a, 16 b).
In figures for convenience we have unrolled the areas, which have been glued
together along a line~$ M_1-M_3=N_0 $, where~$ N_0 $ is a solution of the
equation $D-4\Gamma_1\Gamma_2 N_0=0$.
The difference of a plane from a sphere is exhibited in form of physical areas.
For a plane (see Fig.~16 a) two types of trajectories exist:
\begin{enumerate}
\item Trajectories are touched once boundaries~$\Delta_i=0 $ and then leaving
to infinity;
\item Trajectories stayed between boundaries~$\Delta_i=0 $.
\end{enumerate}
Different motions of vortices correspond to these cases: in the first case the
vortices run up, passing only once through a collinear configuration, in second
case vortex pairs alternately pass one through another ({\it leapfrog of
Helmholtz}), remaining on bounded distance. Only trajectories of the second type
exists for a sphere. The analysis of leapfrog on a plane was carried out by
other method in~\cite{Melesh}.

Let us introduce also, for completeness, the equation of motion and geometrical
interpretation at the zero moment~$ D=0 $ (necessary condition of a
collapse~\cite{NovikovSed}).

According to~(\ref{zerc6}) in this case~$ y=0 $, and all mutual
distances~$M_1, \dots, M_4 $ can be expressed via variables~$ x, M $ by
formulas
$$
\begin{array}{c}
M_2=\frac{\Gamma_2x}{\Gamma_1(\Gamma_1+\Gamma_2)}\,,\quad
M_4=\frac{\Gamma_1x}{\Gamma_2(\Gamma_1+\Gamma_2)},\quad N =\frac{x}{\Gamma_1
+\Gamma_2}\,, \\ [10pt]
M_1 =\frac12 (M+N), \quad M_3 =\frac12 (M-N)\,.
\end{array}
$$
The equations of motion for~$ m = (\Gamma_1 +\Gamma_2) M$ and $x$ have a form
\begin{equation}
\label{zerc10}
\begin{aligned}
\frac{dx}{d\tau}&=-2x^2\,,\\
\frac{dm}{d\tau}&=\frac12\left(\frac{\Gamma_1}{\Gamma_2}+
\frac{\Gamma_2}{\Gamma_1}\right)m^2-2xm-\frac12\left(\frac{\Gamma_1}
{\Gamma_2}+\frac{\Gamma_2}{\Gamma_1}\right)x^2\,.
\end{aligned}
\end{equation}
The trajectory, defined by the system~(\ref{zerc10}), is obtained from equation
\begin{equation}
\label{zerc11}
m=x\frac{Cx ^{4/a} -1}{Cx ^{4/a} -1}\,, \quad a =\frac12\left (\frac{\Gamma_1}
{\Gamma_2}+\frac{\Gamma_2}{\Gamma_1}\right)\,,
\end{equation}
where~$ C $ is a constant of an integration. The equation, defining the form of
area~$ (\Delta_i > 0)$ on a plane~$ m, x $ has a form
\begin{equation}
\label{zerc12}
\Gamma_1\Gamma_2\left(x(m-ax)-\frac{(m^2-x^2)x}{8R^2(\Gamma_1+\Gamma_2)^2}
\right) > 0\,.
\end{equation}
Analyzing~(\ref{zerc11}), (\ref{zerc12}) near to the origin of
coordinates~$m=x=0 $ it is possible to conclude that for reflectional
symmetrical solution the simultaneous collapse of four vortices is impossible.

The discussed above integrable systems of vortex dynamics allow the complete
enough analysis with the help of qualitative research of dynamical systems on a
plane. At the same time application of classical methods of an explicit solution
with the help of the theory of special (Abelian) functions produces very bulky
expressions, and that expressions are not permitting to make any of
representation about the real motion~\cite{Goryachev1}.

\section { Algebraization and reduction of the three-body problem }

The reduction in the three-body problem of  celestial mechanics
is for the first time systematically considered in the lectures of Jacobi
~\cite {Jacobi}. He dwelt upon  barycentric coordinates, which permit
to ignore uniform motion of the centre of masses, and also to      the
procedure of elimination of momentum (elimination of a knot).
Constructively this process reducing in a plane (spatial) case in a system with
three (four) degrees of freedom was executed by Radout, Brunce, Charlier,
Lie, Levi-Chevita and Whittaker, the review of their researches contains in the~\cite
{Whitt, Sharle}. The most natural and complete reduction of order
in the three-body problem belongs to van Kampen (E.\,P.\,van Kampen) and
Wintner (A.\,Wintner)~\cite {Camp}.
This problem is also considered in the work ~\cite {Wint}. All these classical
results are based on the theory of canonical transformations and connected with bulky calculations.
Here we give more geometrical method of the reduction of order in a flat three-body
problem, which could be easily generalized to a spatial~$n$-body problem. It is connected with a preliminary
algebraization of the reduced system and subsequent introduction of canonical
coordinates on a symplectic orbit. In contrast with the classical approach this
procedure uses (algebraic) Poisson structure of the reduced system only and has no connection with Hamiltonian
(and therefore it can be derived for other
potentials, which depend on mutual distance between bodies). The algorithm
of reduction under our consideration is universal enough and uses only algebraic methods.
Undoubdedly, the study of the algebraic form of~$n$-body problem
(not mentioning  more natural symmetrization) has large prospects for celestial mechanics.

\subsection {An algebraization of a system}
The Hamiltonian of a plane three-body problem can be written in the form
\begin {equation}
\label {New1/2}
H = \frac {1} {2} \sum _ {i=1} ^k \frac {{\bf p} _i^2} {m_i} + \frac {1} {2} \sum
_ {i\ne j} ^3 U (| {\bf r} _i- {\bf r} _j |) \,
\end {equation}
where $ {\bf r} _i (x_i, y_i), \, {\bf p} _i (p _ {ix}, p _ {iy}) $ are
two-dimensional vectors of position and momentums of particles, for which
components ($p _ {ix}, \, x_i, \, p _ {iy}, \, y_i $) of Poisson brackets are
canonical.

Let us choose new variables, which characerize relative dynamics of particles
as the quadratic forms by canonical variables:
\begin {enumerate}
\item Quadrates of mutual distances
\begin {equation}
\label {New1}
\begin {aligned}
 M_1&=|{\bf r} _3- {\bf r} _2 | ^ 2, \\
 M_2&=|{\bf r} _1- {\bf r} _3 | ^ 2, \\
 M_3&=|{\bf r} _2- {\bf r} _1 | ^ 2.
\end {aligned}
\end {equation}
\item Scalar products of momentums and neighbouring sides of a triangle
formed by points (see Fig.~17)
\begin {equation}
\label {New2}
\begin {aligned}
 X_2&=({\bf r} _1- {\bf r} _3, {\bf p}_1)\mbox{, }&
 X_3&=({\bf r} _2- {\bf r} _1, {\bf p} _1)\mbox{, }\\
 Y_1&=({\bf r} _3- {\bf r} _2, {\bf p}_2)\mbox{, }&
 Y_3&=({\bf r} _2- {\bf r} _1, {\bf p} _2)\mbox{, }\\
 Z_1&=({\bf r} _3- {\bf r} _2, {\bf p}_3)\mbox{, }&
 Z_2&=({\bf r} _1- {\bf r} _3, {\bf p} _3)\,. \\
\end {aligned}
\end {equation}

\end {enumerate}

Since the Hamiltonian (\ref {New1/2}) is dependent only on absolute
quantity of momentums and mutual distances, it can be expressed through
relative variables (\ref {New1}--\ref {New2}). Moreover these variables
form a Lie-Poisson bracket
\begin {equation}
\label {New3}
\begin {aligned}
\{X_2,X_3\}&=X_2+X_3\mbox{, }& \{X_2,Y_1\}&=0\mbox{, }& \{X_2,Y_3\}&=-Y_1 - Y_3\,, \\
\{X_2,Z_1\}&=X_2+X_3\mbox{, }& \{X_2,Z_2\}&=-X_2-Z_2\mbox{, }& \{X_3,Y_1\}&=-X_2-X_3\,,\\
\{X_3,Y_3\}&=X_3+Y_3\mbox{, }& \{X_3,Z_1\}& = 0\mbox{, }& \{X_3,Z_2\}&=Z_1+Z_2\,, \\
\{Y_1,Y_3\}&=-Y_1-Y_3\mbox{, }& \{Y_1,Z_1\}&=Y_1+Z_1\mbox{, }& \{Y_1,Z_2\}&=-Y_1-Y_3\,,\\
\{Y_3,Z_1\}&=-Z_1-Z_2\mbox{, }& \{Y_3,Z_2\}&= 0\mbox{, }& \{Z_1,Z_2\}&=Z_1+Z_2\,,\\
\{X_2,M_1\}&= 0\mbox{, }& \{X_2,M_2\}&=-2M_2\mbox{, }& \{X_2,M_3\}&=M_1-M_2-M_3\,, \\
\{X_3,M_1\}&=0\mbox{, }& \{X_3,M_2\}&=-M_1+M_2+M_3\mbox{, }& \{X_3,M_3\}&=2M_3\,,\\
\{Y_1,M_1\}&=2M_1\mbox{, }& \{Y_1,M_2\}&=0\mbox{, } &\{Y_1,M_3\}&=-M_2+M_1+M_3\,,\\
\{Y_3,M_1\}&=M_2-M_1-M_3\mbox{, } &\{Y_3,M_2\}&=0\mbox{, }& \{Y_3,M_3\}&=-2M_3\,,\\
\{Z_1,M_1\}&=-2M_1\mbox{, }& \{Z_1,M_2\}&=M_3-M_1-M_2\mbox{, }& \{Z_1,M_3\}&=0\,, \\
\{Z_2,M_1\}&=-M_3+M_1+M_2\mbox{, }& \{Z_2,M_2\}&=2M_2\mbox{, }& \{Z_2,M_3\}& = 0\,, \\
\{M_1,M_2\}&=0\mbox{, }& \{M_3,M_2\}&= 0\mbox{, }&\{M_1,M_3\}& = 0\,.
\end {aligned}
\end {equation}

It can be shown, that variables $M, X, Y, Z $ commute with an integral of the total angular momentum of a system with respect to an arbitrary point and quadrate of total momentum. Hence, map\-ping~{(\ref {New1}--\ref {New2})} corresponds to the
reduction with these integrals, while the rank of an initial canonical bracket
(equal 12) reduces four units down.
Thus rank of a bracket (\ref {New3}) is equal to eight, and the Casimir function
is the quadrate of total momentum (which is expressed through relative variables (\ref {New1}--\ref {New2}) in contrast to the angular momentum).

{\bf Remark 10.} {\it The method of the algebraization of a dynamical
system which was used here and in the vortex dynamics is a special case of
a general method. It is based on the fact that the quadratic functions of
variables $p, x $ form the algebra~ $ sp (n) $ with respect to canonical
symplectic structure~ $\omega =\sum _ {i} dp_i\wedge dx_i $ ~ \cite
{ArnoldGivental}}. {\it There exists a certain subalgebra in ~ $ sp (n) $,
and its generators commute with integrals of motion of a system to each
particular problem (Hamiltonian).}

\subsection {Barycentric coordinates and Poisson
submanifolds}
The above mentioned reduction concerns an arbitrary inertial system,
for which two projections of total momentum and total angular
momentum form in general case a noncommutative set of integrals.
In the system of the centre of inertia (barycentric frame) the total momentum is equal to zero, hence set of integrals is commutative and the reduction for one more degree of freedom is possible. To reduce the order in the algebraic form it is  necessary to restrict a system on a submanifold of zero total
momentum in the algebra (\ref{New3}).

We choose new generators of the algebra~(\ref {New3}).
They correspond to eigenvectors of Killing form~\cite {Barut}
\begin {equation}
\label {New4}
\begin {gathered}
\begin {aligned}
S_1 &= \frac{2}{\sqrt{3}}(X_2+X_3+Y_1+Y_3+Z_1+Z_2)\,,\qquad &
S_2 &= \frac {1} {4} (Y_3+Z_2-X_2-Y_1) \,, \\
S_3 &= \frac {1} {4\sqrt {3}} (2X_3-Y_3+2Z_1-Z_2+X_2-Y_1) \,, \qquad &
S_4 &= \frac {1} {6} (X_2-X_3-Y_1+Y_3+Z_1-Z_2) \,, \\
S_5 &= X_2 - Y_1 - Y_3 + Z_2 \,,\qquad &
S_6 &= - X_2 - X_3 + Y_1 + Z_1 \,, \\
\end {aligned} \\
N_1 = -\frac {1} {\sqrt {6}} (M_1+M_2+M_3) \,, \qquad
N_2 = \frac {1} {\sqrt {2}} (M_1 - M_3) \,, \qquad
N_3 = -\frac {1} {\sqrt {6}} (M_1 -2M_2 + M_3) \,.
\end {gathered}
\end {equation}

The variables $S_5, S_6 $ are proportional to projections of total momentum
on two sides of a triangle (see Fig.~17). A linear span~$ S_5, S_6 $
forms an ideal of algebra (\ref {New3}), hence submanifold of zero
momentum
\begin {equation}
\label {New4.5}
S_5=0 \,,\qquad S_6=0
\end {equation}
is Poisson submanifold. Restricting a system on~(\ref {New4.5}) and ordering
remaining variables as follows~$ {\bf x} = (S_1, \, S_2, \, S_3, \, S_4,\,N_1, \, N_2, \, N_3) $, we
obtain the table of commutators
\begin {equation}
\label {New5}
\| \{x_i, x_j \} \| =
\left (\begin {array} {ccccccc}
0 & S_3 & -S2 & 0 & 0 & N_3 & -N_2 \\
-S_3 & 0 & -S_1 & 0 & -N_3 & 0 & - N_1 \\
S2 & S_1 & 0 & 0 & N_2 & N_1 & 0 \\
0 & 0 & 0 & 0 & -N_1 & -N_2 & -N_3 \\
0 & N_3 & -N_2 & N_1 & 0 & 0 & 0 \\
-N_3 & 0 & -N_1 & N_2 & 0 & 0 & 0 \\
N_2 & N_1 & 0 & N_3 & 0 & 0 & 0
\end {array} \right).
\end {equation}
The Casimir function of structure (\ref {New5}) coincides with quadrate of
the total angular momentum with respect to the centre of masses
\begin {equation}
\label {New6}
M_z^2=4\frac {\left < {\bf S, N} \right > ^2} {\left < {\bf N, N} \right >} \,
\end {equation}
where~$ \left < {\bf a}, {\bf b} \right > =a_1b_1-a_2b_2-a_3b_3 $ is a scalar
product in the Minkowski space.
Numerator and denominator of this fraction are Casimir functions of a
subalgebra~$ so (1,2) \otimes_s\mathbb{R}^3 $ with generators
$ (S_1, S_2, S_3, N_1, N_2, N_3) $.
Moreover~$ \left < {\bf N}, {\bf N} \right > =2\Delta^2 $, where $ \Delta $  is
the square of a triangle formed by particles.

The algebra ~ $ l_7 $, corresponding to a bracket (\ref {New5}) is the
semidirect sum of subalgebra formed by elements $ {S_i, \, i=1,
\ldots, 4} $ and three-dimensional commutative ideal: $l_7 = (so (1,2) \oplus
R) \oplus_s R^3 $.

The quadrates of momentum and mutual distances can be written as
\begin{equation}
\label {New7}
\begin {gathered}
p_k^2 = \frac {1} {3} \frac {\left < {\bf e} _k, {\bf N} \right > S_4^2
- 2\left < {\bf e} _k, {\bf S}, {\bf N} \right > S_4 +
2\left < {\bf e} _k, {\bf S} \right > \left < {\bf S}, {\bf N} \right > \left <
{\bf e} _k, {\bf N} \right > \left < {\bf
S}, {\bf S} \right >} {\left < {\bf N}, {\bf N} \right >} \, \\
M_k =\langle {\bf e} _k, {\bf N} \rangle \mbox{, } k=1, \, 2, \, 3,
\end {gathered}
\end {equation}
where
$ {\bf S} = (S_1, S_2, S_3), \, {\bf N} = (N_1, N_2, N_3) $ and vectors $ {\bf e}
_k $ have the form
$$
{\bf e}_1=\frac{1}{\sqrt{6}}(-2,-\sqrt{3},1)\mbox{, } {\bf e} _2 =\frac
{1} {\sqrt {6}} (-2,0, -2)\mbox{, } {\bf
e} _3 =\frac {1} {\sqrt {6}} (-2, -\sqrt {3}, 1)
$$
and satisfy the relation
$ \langle {\bf e} _i, {\bf e} _j\rangle=1-\delta _ {ij} $. Here~$ \langle {\bf a},
{\bf b}, {\bf c} \rangle $ is a determinant matrix which is formed by components of
vectors
$ \bf a, b, c $.

The Hamiltonian has the form
\begin {equation}
\label {New9}
H = \frac {p_1^2} {m_1} + \frac {p_2^2} {m_2} + \frac {p_3^2} {m_3} + \frac
{m_1m_2} {\sqrt {M_3}} + \frac {m_2m_3} {\sqrt {M_1}} + \frac {m_1m_3} {\sqrt{M_2}}\,.
\end {equation}

\subsection {Orbits and symplectic coordinates} In algebra (\ref {New5}) the
orbits appropriate to physical motions of a system are divided in
two types.
\begin {enumerate}

\item Regular six-dimentional orbits are surfaces of a level of Casimir
function~(\ref {New6}). The quadrate of area of a triangle is nonnegative,
therefore for physical symplectic sheets~$ \left < {\bf N, N} \right > \ge 0 $.
 Topologi\-cal\-ly such orbit is diffeomorphic~$TL^2\times \mathbb{R} ^
{+}\times \mathbb{R} $. Here $TL^2 $ ~ is a tangent bundle of a
pseudosphere $ \left < {\bf N, N} \right > =c_1 $,  its ``radius" has
non-negative values $c_1\in \mathbb{R} ^ {+} = [0, \infty) $, and the last factor
corresponds to a linear span $S_4 $.

\item It is possible to show (analyzing a rank of a matrix (\ref {New5})),
that four-dimentional orbits diffeomor\-phic to tangent bundle to
a cone $ \left < {\bf N}, {\bf N} \right > =0 $ pass through points satisfying
the equations ${ \left < {\bf N}, {\bf N}
\right > =0}, \left < {\bf S}, {\bf N} \right > =0 $. These orbits correspond to
the motion of particles along a straight line (the area~$\Delta$ is equal to zero).
\end {enumerate}

{\bf Remark 11.} {\it
The orbits passing through points with negative value of
quadrate of area~$ \left < {\bf N}, {\bf N} \right > < 0 $ or with zero
quadrates of distances have no physical sense.}

We construct symplectic coordinates on a regular orbit similar to
Anduier-Depri coordinates in the rigid body dynamics~\cite {BE}.
For this purpose we will apply the following algorithm.

For algebra (\ref {New5}) we shall consider the following consequence of the
inserted subalgebras $so (1,2) \subset so (1,2) \oplus_s R^3\subset l_7 $.

On a subalgebra $so (1,2) $ we shall choose $L=S_1 $ as an action.
For a Hamiltonian field of vectors with the Hamilton function~$ \mathcal H=L $
$$
\frac {dS_1} {dl} =0 \,,\qquad \frac {dS_2} {dl} =S_3 \,,\qquad \frac {dS_3} {dl}
= -S_2\,
$$
time parameter along an integral curve $l $ is a desired angular variable
canonically conjugate $L $: ${\{l, L \}=1}$.
Choosing constants of an integration so that the relation
$ \{S_2, S_3 \} =-S_1 $ can be fulfilled, we obtain
\begin {equation}
\label {New10}
S_1=L \,,\qquad S_2 =\sqrt {L^2-G^2} \cos l \,, \qquad S_3 =\sqrt {L^2-G^2}
\sin l \,.
\end {equation}

We shall choose the Casimir function $G =\sqrt {S_1^2-S_2^2-S_3^2} $ of subalgebra
$so (1,2) $ as a Hamiltonian $ \mathcal H=G $ on a subalgebra
$so (1,2) \oplus_s R^3 $. Then we integrate the equation that is linear
with respect to~$N_i $
$$
\frac {dN_1} {dg} = \frac {S_3 N_2 - S_2 N_3} {G}\,, \quad
\frac {dN_2} {dg} = \frac {S_3 N_1 - S_1 N_3} {G}\,, \quad
\frac {dN_3} {dg} = \frac {-S_2 N_1 + S_1 N_2} {G}\,,
$$
{\sloppy We find dependence of constants of integration on $L, l, G $  from
commutators of the subalgebra ${so (1,2) \oplus_s R^3}$.
\begin{equation}
\label {New11}
\begin {array} {l}
N_1 =\frac {HL} {G^2}
+\frac{\sqrt{L^2-G^2}}{G}\sqrt{\frac{H^2}{G^2}- \Delta^2} \cos g\,, \\
N_2 =\frac {H} {G^2} \sqrt {L^2 - G^2} \cos l
+\frac{L}{G}\sqrt{\frac{H^2}{G^2}-\Delta^2}\cos g\cos l - \sqrt {\frac {H^2}
{G^2} - \Delta^2} \sin g\sin l\,, \\
N_3 =\frac {H} {G^2} \sqrt {L^2 - G^2}
\sin l +\frac{L}{G}\sqrt{\frac{H^2}{G^2}-\Delta^2}\cos g\sin l + \sqrt
{\frac {H^2} {G^2} - \Delta^2} \sin g\cos l\,,
\end {array}
\end {equation}
where $H=\left<{\bf S},{\bf N}\right>,\;\Delta=\left<{\bf N},{\bf N}\right>$.

}

We shall choose as a last action variable~$S=S_4 $.
As a result we obtain the following expressions
\begin {equation}
\label {New12}
\begin {aligned}
 N_1&=e^s\Bigl(\frac{M_z L} {G^2}
+\frac{\sqrt{L^2\,{-}\,G^2}}{G}\sqrt{\frac{M_z^2}{G^2}-1}
\cos g\Bigr), \\
 N_2&=e^s\Bigl(\frac{M_z}{G^2}\sqrt{L^2{-}G^2}\cos l
+\frac{L}{G}\sqrt{\frac{M_z^2}{G^2}-1}\cos g\cos l - \sqrt {\frac {M_z^2}
{G^2} - 1} \sin g\sin l \Bigr), \\
N_3&=e^s\Bigl(\frac{M_z}{G^2}\sqrt{L^2-G^2}\sin l
+\frac{L}{G}\sqrt{\frac{M_z^2}{G^2}-1}\cos g\sin l + \sqrt {\frac {M_z^2}
{G^2} - 1} \sin g\cos l \Bigr), \\
 S_4&=S\,,
\end {aligned}
\end {equation}
here $M_z $ is a constant of total momentum with respect to a system of the
centre of masses. The variables $S_1, S_2, S_3 $ express by the formulas
(\ref {New10}).

It is possible to introduce the other pair of canonical variables~$x,\,y $,
instead of~$s,\,S $, by the formulas~$ e^s=x, \, S=xy $.

{\bf Remark 12.} {\it Using the above mentioned algebraic structure of
three-body problem, it is also easy to reconstruct the results of classics
in a problem of reduction of order. However, obtained expressions will
remain bulky enough. Remind that in researches of Brunce, Whittaker and
van Kampen as positional variables the mutual distances are chosen.
Reduction of the order which has been carried out by Charlier, uses
distances of three bodies from the general centre of inertia.}

Using the algebraic form of the equations of the three-body problem
it is easy to investigate its particular solutions. One of such
solutions was founded by Lagrange. In this case all of three bodies are
situated in tops of an equilateral triangle (which in an even more special
case does not change sizes). The second solution belongs to Euler and
defines collinear stationary configurations. All these solutions can be
defined by a method, explained in \cite {Wint, Smeil}, and also one can
obtain a system of appropriate invariant relations which are not determined,
Poisson manifolds.

It is easy to show (see ~\cite {Wint}), that there are only two ``rigid-body" stationary configurations in the three-body problem --- Euler and
Lagrangian. Collinear configurations are investigated in detail in the~ $ n $-body problem. By the Multon theorem their number is equal
to~$\frac {n!} {2} $~\cite {Smeil}.

The topological proof of the Multon theorem is given with the use of the Morse theory.
In the paper~\cite {Smeil} series of hypotheses about the amount of noncollinear configurations in a~$ n $-body
problem are also expressed. As far as we know the majority of them are not proved until now. We hope that the algebraic form of the reduced system offered in this application, will allow to achieve further progress in this problem.

The autors are grateful to N.\,N.\,Simakov for the help in numerical computations. In Russian outcomes
of this paper are introduced in the book A.\,V.\,Borisov,
I.\,S.\,Mamaev ``Poisson structures and Lie algebras in  Hamiltonian
mechanics".

\begin{figure}[ht!]
$$
\includegraphics{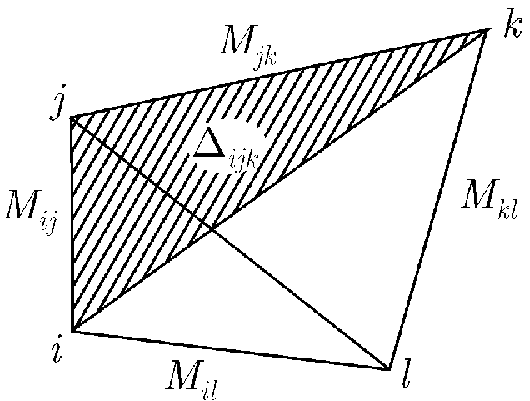}
$$
\caption{}
\end{figure}

\begin{figure}[ht!]
$$
\includegraphics{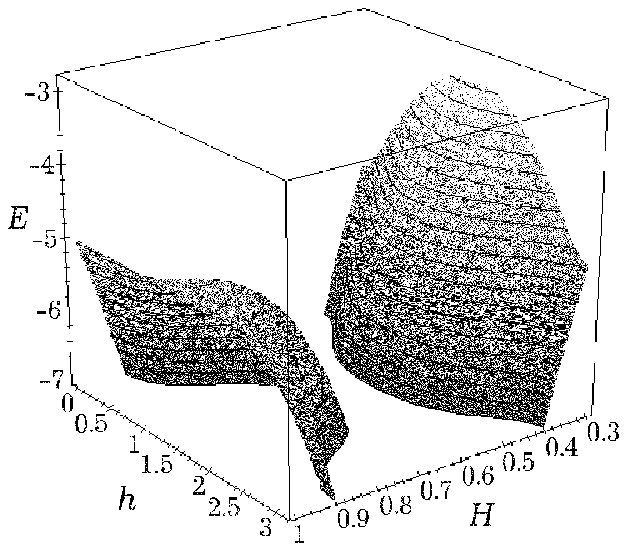}
$$
\caption{$G=0.3,\;g=\pi/2$}
\end{figure}

\begin{figure}[ht!]
$$
\includegraphics{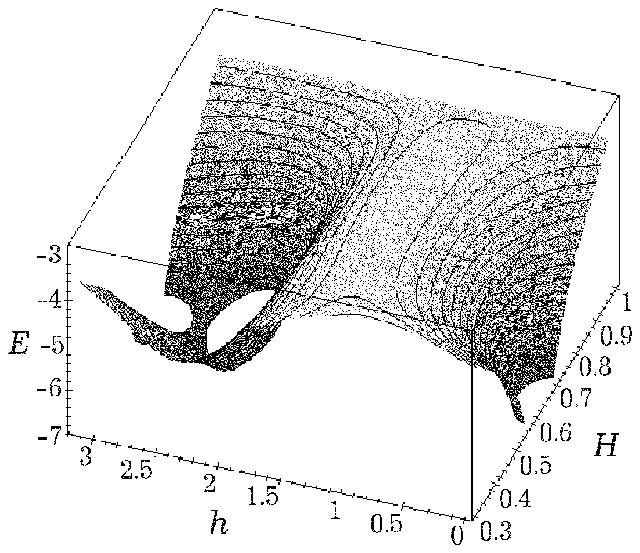}
$$
\caption{$G=0.3,\;g=\pi/4$}
\end{figure}

\begin{figure}[ht!]
$$
\includegraphics{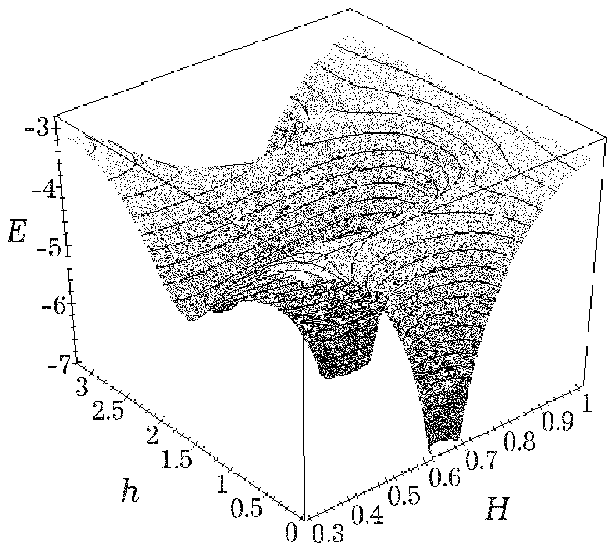}
$$
\caption{$G=0.3,\;g=0$}
\end{figure}

\begin{figure}[ht!]
$$
\includegraphics{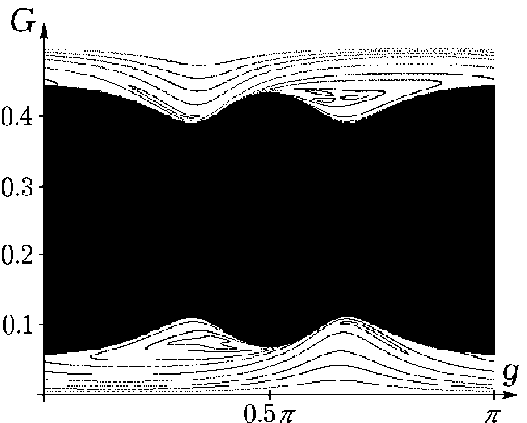}
$$
\caption{$H=0.5,\;E=-3.2$}
\end{figure}

\begin{figure}[ht!]
$$
\includegraphics{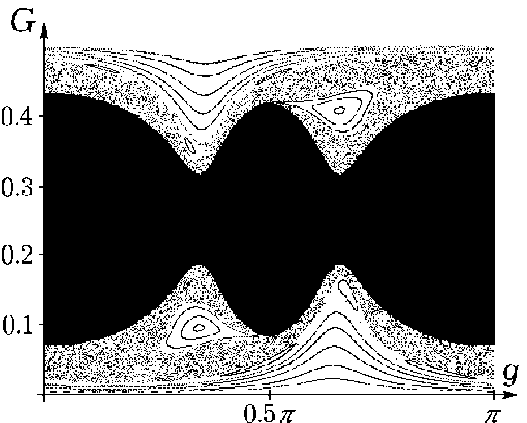}
$$
\caption{$H=0.5,\;E=-3.3$}
\end{figure}

\begin{figure}[ht!]
$$
\includegraphics{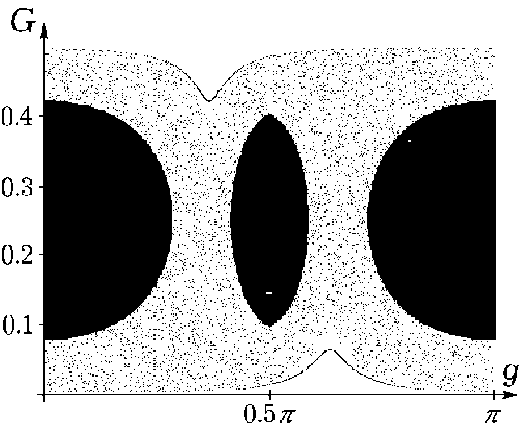}
$$
\caption{$H=0.5,\;E=-3.4$}
\end{figure}

\begin{figure}[ht!]
$$
\includegraphics{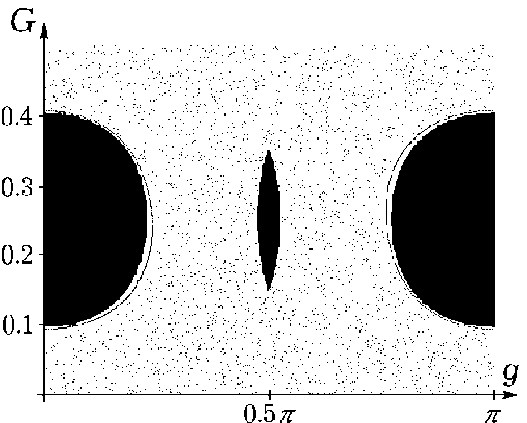}
$$
\caption{$H=0.5,\;E=-3.6$}
\end{figure}

\begin{figure}[ht!]
$$
\includegraphics{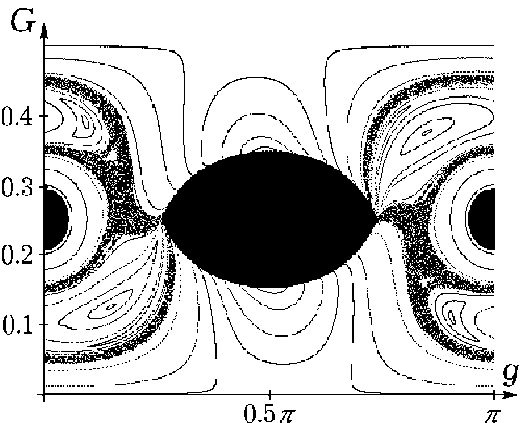}
$$
\caption{$H=0.5,\;E=-6.0$}
\end{figure}

\begin{figure}[ht!]
$$
\includegraphics{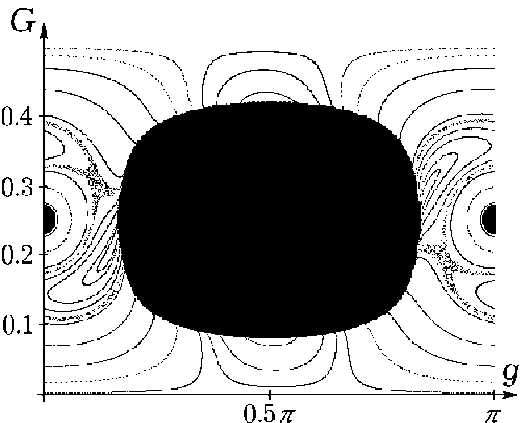}
$$
\caption{$H=0.5,\;E=-7.5$}
\end{figure}

\begin{figure}[ht!]
$$
\includegraphics{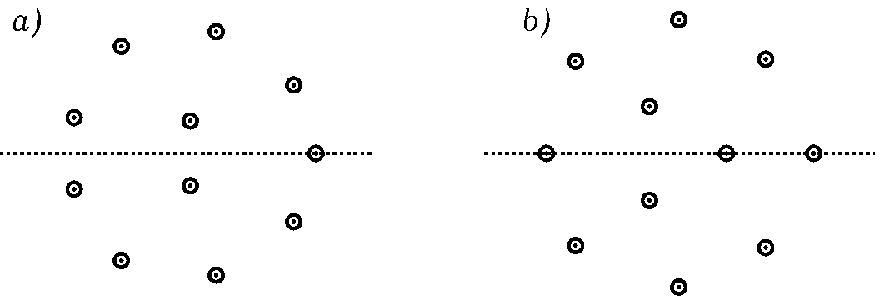}
$$
\caption{}
\end{figure}

\begin{figure}[ht!]
$$
\includegraphics{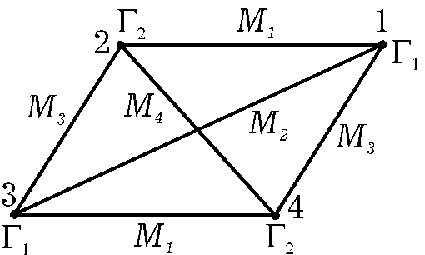}
$$
\caption{}
\end{figure}

\begin{figure}[ht!]
\begin{center}
{ \small
\begin{tabular}{ccc}
\includegraphics{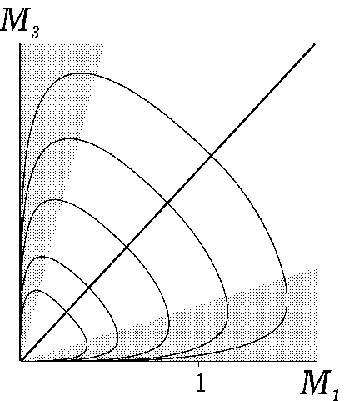} & \includegraphics{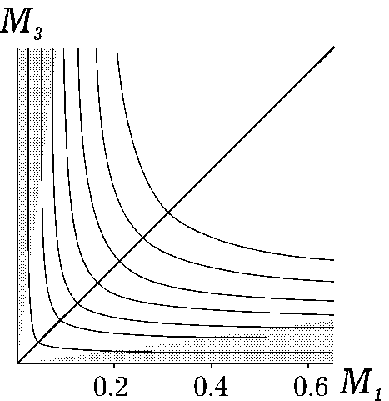} &
\includegraphics{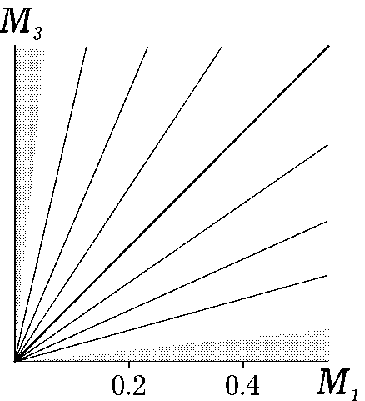}\\
a) $\begin{aligned}\Gamma_1&=-1.5,\\
 \Gamma_2&=2.0.
 \end{aligned}$&
b) $\begin{aligned}\Gamma_1&=-0.183,\\
 \Gamma_2&=2.0. \end{aligned}$&
c) $\begin{aligned}\Gamma_1&=-0.183,\\
 \Gamma_2&=0.683.\end{aligned}$
\end{tabular}}
\end{center}
\caption{}
\end{figure}

\begin{figure}[ht!]
\begin{center}
{ \small
\begin{tabular}{ccc}
 \includegraphics{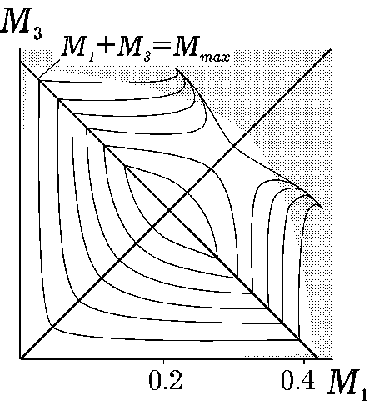} & \includegraphics{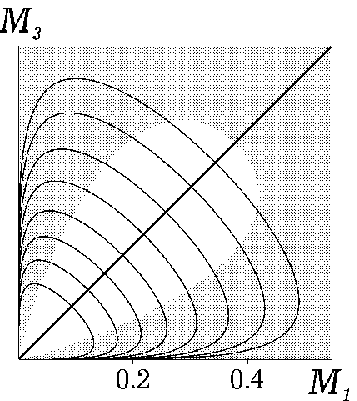} &
\includegraphics{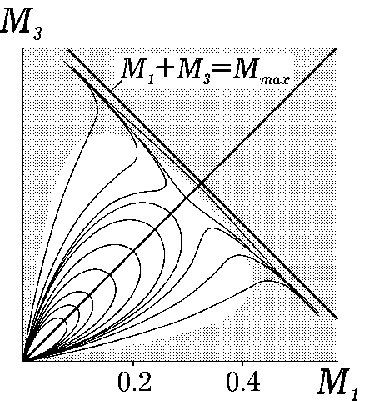}\\
a) $\begin{aligned}\Gamma_1&=-0.183,\\
 \Gamma_2&=2.0,\\
 k&=0.6.\end{aligned}$&
b) $\begin{aligned}\Gamma_1&=-1.5,\\
 \Gamma_2&=2.0,\\ k&=0.6.\end{aligned}$&
c) $\begin{aligned}\Gamma_1&=-0.183,\\
 \Gamma_2&=0.683,\\
 k&=0.6.\end{aligned}$
\end{tabular}} \vspace{-5mm}
\end{center}
\caption{}
\end{figure}

\begin{figure}[ht!]
$$
\includegraphics{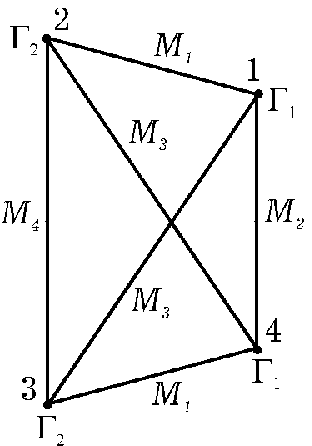}
$$
\caption{}
\end{figure}

\begin{figure}[ht!]
\begin{center}
{ \small
\begin{tabular}{cc}
\includegraphics {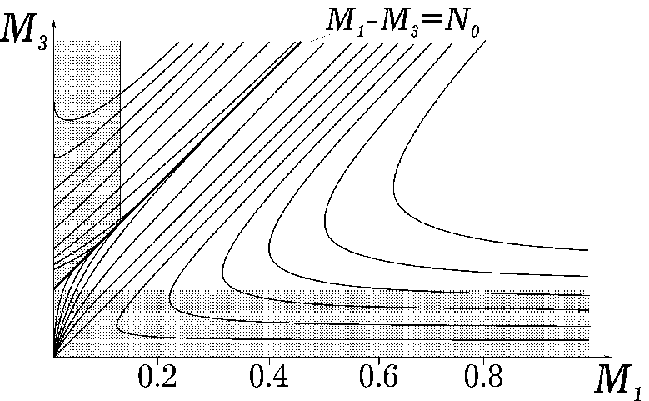} & \includegraphics{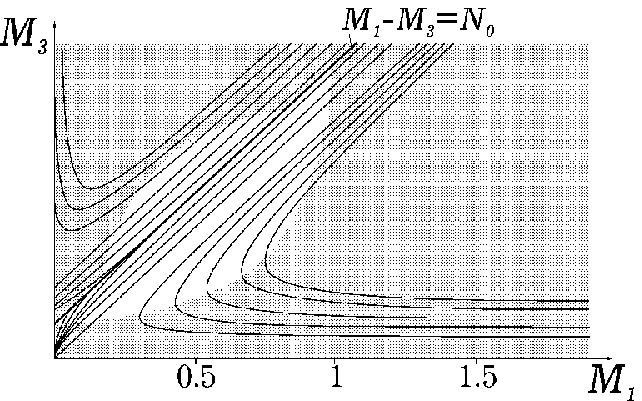}\\
 $\begin{aligned}\Gamma_1&=2.0,\;&\Gamma_2&=2.0\\
k&=0.0,\;&D&=-2.0.\end{aligned}$&
$\begin{aligned}\Gamma_1&=2.0,\;&\Gamma_2&=2.0\\
k&=0.9,\;&D&=-2.0.\end{aligned}$\\[15pt]
\includegraphics {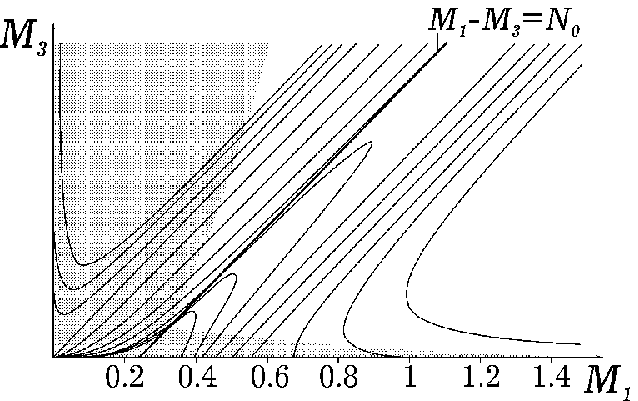} & \includegraphics {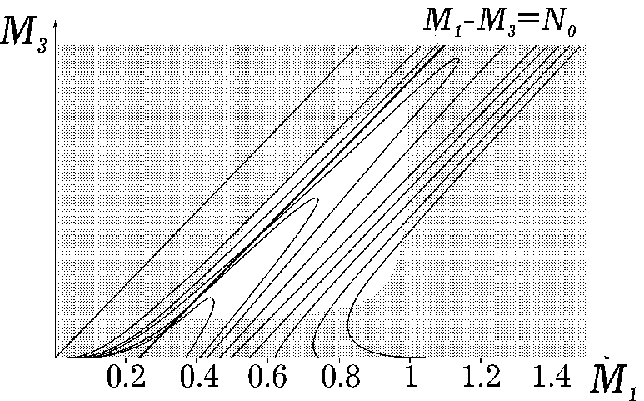}\\
 $\begin{aligned}\Gamma_1&=-1.0,\;&\Gamma_2&=2.0\\
k&=0.0,\;&D&=-2.0.\end{aligned}$&
$\begin{aligned}\Gamma_1&=-1.0,\;&\Gamma_2&=2.0\\
k&=0.9,\;&D&=-2.0.\end{aligned}$\\[15pt]
\includegraphics{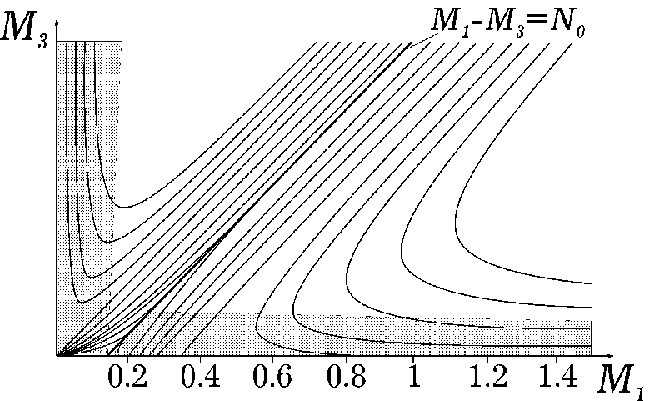} & \includegraphics {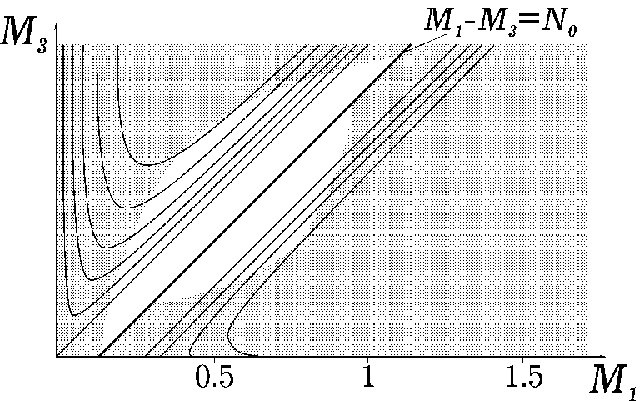}\\
 $\begin{aligned}\Gamma_1&=-1.8,\;&\Gamma_2&=2.0\\
k&=0.0,\;&D&=-2.0.\end{aligned}$&
$\begin{aligned}\Gamma_1&=-1.8,\;&\Gamma_2&=2.0\\
k&=0.9,\;&D&=-2.0.\end{aligned}$\\[15pt]
 a)&b)
\end{tabular}} \vspace{-5mm}
\end{center}
\caption{}
\end{figure}

\begin{figure}[ht!]
$$
\includegraphics{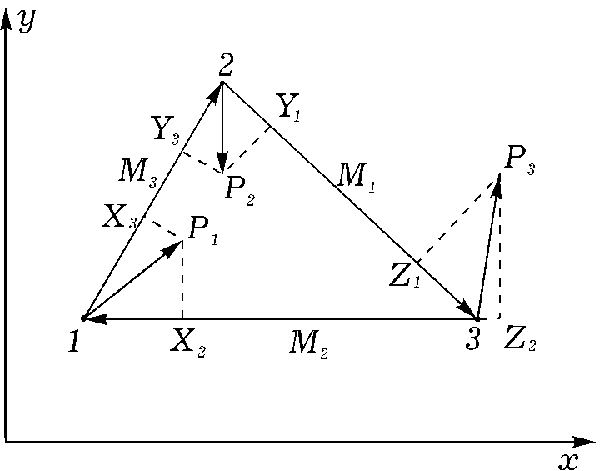}
$$
\caption{}
\end{figure}

\end{document}